\documentclass[a4paper, final]{siamart251104}

\usepackage{graphicx}%
\graphicspath{{./Figures/}}

\usepackage{multirow}%
\usepackage{amsmath,amssymb,amsfonts}%
\usepackage{mathrsfs}%
\usepackage[title]{appendix}%
\usepackage{xcolor}%
\usepackage{textcomp}%
\usepackage{manyfoot}%
\usepackage{booktabs}%
\usepackage{tcolorbox}
\usepackage[capitalize,nameinlink]{cleveref}

\crefname{equation}{}{}
\crefname{section}{section}{sections}
\crefname{subsection}{subsection}{subsections}
\crefname{figure}{Figure}{Figures}
\Crefname{equation}{Equation}{Equations}
\Crefname{section}{Section}{Sections}
\Crefname{subsection}{Subsection}{Subsections}
\Crefname{figure}{Figure}{Figures}
\crefname{appendix}{appendix}{appendices}
\Crefname{appendix}{Appendix}{Appendices}

\usepackage{listings}%
\usepackage{caption}
\usepackage{comment}
\usepackage{subcaption}
\usepackage{rotating} 
\usepackage{algorithm}
\usepackage{algorithmic}

\usepackage{todonotes}
\newcommand{\epsi}{\varepsilon}

\DeclareMathOperator{\Id}{Id}

\DeclareMathOperator{\rmse}{RMSE}

\Crefname{algocf}{Algorithm}{Algorithms}

\newcommand{\Dx}{{\Delta r}}
\newcommand{\Dt}{{\Delta t}}
\newcommand{\E}{\mathbb{E}}

\newcommand{\F}{\mathcal{F}}

\newcommand{\U}{\mathcal{U}}

\newcommand{\R}{\mathbb{R}}
\newcommand{\N}{\mathbb{N}}
\newcommand{\RR}{\mathbb{R}^2}
\newcommand{\RSet}{\mathbb{R}}

\newcommand{\Meff}{\ensuremath{M_{\textrm{eff}}}}

\def\mathbi#1{\textbf{\em #1}}

\title{State and Parameter Estimation for a Neural Model of Local Field Potentials}

\author{%
  Daniele Avitabile%
  \thanks{%
    Amsterdam Centre for Dynamics and Computation,
    Vrije Universiteit Amsterdam,
    Department of Mathematics,
    Faculteit der Exacte Wetenschappen,
    De Boelelaan 1081a,
    1081 HV Amsterdam, The Netherlands.
  \protect
    MathNeuro Team,
    Inria branch of the University of Montpellier,
    860 rue Saint-Priest
    34095 Montpellier Cedex 5
    France.
  \protect
  (\email{d.avitabile@vu.nl}, \url{www.danieleavitabile.com}, \url{www.amsterdam-dynamics.nl}).
  }
  \and
  Gabriel J. Lord
  \thanks{
    Radboud University, Nijmegen, The Netherlands.
    \protect(\email{gabriel.lord@ru.nl})
  }
  \and
  Khadija Meddouni
  \thanks{Radboud University, Nijmegen, The Netherlands.
    \protect(\email{khadija.meddouni@ru.nl})
  }
}

\begin{document}

\maketitle
\begin{abstract}
The study of cortical dynamics during different states such as decision making, sleep
and movement, is an important topic in Neuroscience. Modelling efforts aim to relate
the neural rhythms present in cortical recordings to the underlying dynamics
responsible for their emergence. We present an effort {\color{black}to capture certain features of the neural
activity} from the cortex of a mouse during natural sleep, obtained through local
field potential measurements. Our approach relies on using a discretized
Wilson--Cowan Amari neural field model for neural activity, along with a data
assimilation method that allows the Bayesian joint estimation of the state and
parameters. We demonstrate the feasibility of our approach on synthetic measurements
before applying it to a dataset available in literature. {\color{black} Our findings suggest the
potential of our approach to characterize the parameters of the external forcing in the neural activity model, while simultaneously inferring
a state that aligns with the observed signal.}
\end{abstract}

\section{Introduction}\label{section:intro}
Data assimilation is a form of inverse problem that enhances model prediction using
partial and noisy observations, and in which the unknown quantity is a
dynamical system's state, its control parameters, or a combination of these
quantities. 
This approach to model forecast, parameter inference, and uncertainty quantification
has been the subject of an intense work over the past decade, ranging from
theoretical foundations to domain-specific studies in engineering, physics, and
biology \cite{smithUncertaintyQuantificationTheory2014, 
 Law2015,
nakamuraInverseModelingIntroduction2015,
Reich2015,
Sanz-alonsoInverseProblemsData2023}. 

The present paper tests and showcases the data assimilation methodology in the
context of brain dynamics. {\color{black}Our long-term motivation is to develop a numerical
  tool to gain a better understanding of how cortical and limbic brain regions
  interact during sleep}, when memories are thought to consolidate
  \cite{sleep_cortex}. This is a
compelling application for data assimilation, because experimental observations of
brain activity are available for the cortex \cite{pedrosaHippocampalGammaSharp2022},
and one is interested in
describing the coupling with hidden limbic regions. Thus it is interesting to model such interactions, even in a simple
form, and test them against the partial observations of the system, in the upper
(cortical) layers.

Deterministic inverse problems in which the unknown quantity is a resting state or a
static parameter typically suffer from ill-posedness
\cite{vogel_computational_2002,Kirsch_2011}, which can be resolved by appealing to
regularization techniques (see for example
\cite{willoughby_solutions_1979,Kirsch_2011}) or to a Bayesian formulation as
described in
\cite{FRANKLIN1970682,Stuart_2010,Kaipio:1338003,Dashti2017}.

Within the data assimilation framework, in which data is time-dependent, we
distinguish between two categories: methods such as 4D-Var
where the state is estimated over a whole time window of observations, and sequential
methods such as Kalman or particle filters which incorporate the observations
sequentially. The latter is the approach we follow in this paper.

Our aim is to examine Bayesian joint estimation of
parameters and prediction of state from the local field potential measurements. 
Our approach relies on the use of the nested particle filter from
\cite{Crisan2013NestedPF,Miguez23} on a nonlinear state-space system originating from
the discretized deterministic Wilson--Cowan--Amari neural field originally proposed
in
\cite{wilson1973mathematical, Amari}. This equation models heuristically
spatially-dependent brain activity in a cortical domain using coarse activity
variables, their nonlocal synaptic interaction, and \textcolor{black}{the presence of an external
forcing, which may come from other (cortical or non-cortical) regions, from an
external stimulation, or compensate for missing biological terms. We henceforth refer to any of these sources as \textit{external
forcing}, and we stress that the model employed here is not sufficiently granular to
distinguish between them. The present paper focuses on the numerical aspects of the
inversion process, and our model choice has been dictated by mathematical structure
and wide applicability, rather than biological accuracy.}

Although the dynamics of this model have been studied extensively
\cite{ermentrout_mathematical_1979,bressloff_spatiotemporal_2011,coombes_waves_2005,kilpatrick_effects_2010},
there are far fewer works that tackle the inverse problem of estimating
state or model parameters \cite{Schiff_Sauer_08,Potthast,Alswaihli}. In the
context of joint parameter and state estimation, related works include
\cite{KULIKOVA2023104010}, in which the authors derived a continuous-discrete
extended Kalman filter for simultaneous state and parameter estimation for stochastic
dynamic neural field models and demonstrated the robustness of their approach on
synthetic measurements. The authors of \cite{Schiff_Sauer_08} used the unscented Kalman filter to
estimate both state and parameters of a spatiotemporal excitable cortical model and
showed its ability to track spiral waves. In a subsequent work,
\cite{Schiff_Sauer_09} introduced the consensus set method with an ensemble Kalman
filter applied to a homogeneous Wilson-Cowan model to effectively track the state and
infer parameters in both synthetic and experimental measurements.
Recently, state and parameter estimation for the model has been studied from a
theoretical viewpoint and with synthetic data in \cite{piepersethmacher2025guided},
in which data assimilation is introduced with a form of nudging, called guided
process.

The present paper differs from existing literature as it uses experimental
observations with a nonlinear filter, which does not resort to linearised dynamics.
Further, we test the performance of the spatio-temporal
reconstruction against the dataset, and use synthetic data to show how the
methodology accounts for epistemic uncertainty (model imperfections). While this
paper is concerned mostly with the numerics of data assimilation, {\color{black} our
  results show the potential of this framework to to quantify the influence of external forcing
on the modelled cortical activity}, while matching faithfully its experimental
observations.

The paper is organized as follows. In
\Cref{section:model} we present the setting for the estimation problem of interest
including a description of the cortical model and a background on Bayesian
estimation. \Cref{section:synthetic} provides the results of this approach on
synthetic data. We present the experimental cortical data in \Cref{section:data} and
the estimation results in \Cref{section:results}. We conclude in
\Cref{section:conclusion} with a summary of this work, and we detail some additional
computational experiments in the Appendices.

 \section{Problem Setting and Background}\label{section:model}
 
 \subsection{The Neural Field Model}
 We study the data assimilation problem on a discrete version of a one-population Amari neural field model 
\begin{equation}\label{eq:CNF}
    \tau\frac{\partial u(r,t)}{\partial t}=-u(r,t)+\int_D w(r,s)\sigma(u(s,t))ds
    +\Xi(r,t;\theta,a), \qquad r \in D, \qquad t \geq 0
\end{equation}
with initial condition $u(r,0)=u_0(r)$, posed on a one-dimensional, bounded cortical domain $D
\subset \RSet$. The model describes the evolution of the population's neural activity
$u$ at cortical position $r$ and time $t$. The population has nonlinear firing rate
$\sigma$, and typical response time $\tau$, while $w(r,s)$ models synaptic
connections from point $s$ to point $r$ in the cortex. Further, the population is
subject to a spatio-temporal external forcing $\Xi$, which depends upon a set of control
parameters $\theta \in \RSet^{d_\theta}$ that we wish to infer. The forcing may also possibly depend on other parameters $a$ that we keep fixed.
We note that the functions $w$ and $\sigma$ also depend on the parameters, but we omit this dependence for the sake of
conciseness.

We consider synaptic connections described by a Mexican hat kernel, which can model
short-range excitation and long-range inhibition,
\begin{equation}\label{eq:w}
    w(r,s)=A_{ex}\exp\left({\frac{-\Vert r-s
    \Vert^2}{\ell_{ex}^2}}\right)-A_{in}\exp\left({\frac{-\Vert r-s
\Vert^2}{\ell_{in}^2}}\right),
\end{equation}
and a sigmoidal firing rate function with threshold $\mu$ and slope $\lambda$
\begin{equation}\label{eq:f}
    \sigma(u)=\frac{1}{1+\exp(-\lambda(u-\mu))}.
\end{equation}

In this paper, {\color{black} our aim is to characterize
the external forcing $\Xi$} by employing a sequential data assimilation method. To this end,
we fix $\tau$ and all the parameters appearing in $w$ and $\sigma$ to the values given
in \cref{table:params}; we are interested in inferring parameters for the forcing
$\Xi$ and simultaneously estimating a discrete approximation of the state variable $u(\cdot,t)$ from noisy observations. 

Strictly speaking, the model we use is a spatio-temporal discretisation of the
continuum model \cref{eq:CNF}, for specific choices of the spatio-temporal stimulus $\Xi$ which are
described later, in \cref{section:synthetic,section:results}, as they change
depending on the setup of the problem and numerical experiments. We highlight,
however, that we do not quantify how spatial and temporal discretizations affect
the data assimilation process, but we rather take the discrete model as the starting
point for our investigation. The rest of this section is devoted to describing this discrete model.

We take $D = [0,L]$ and discretize space using $J$ evenly spaced nodes $r_0,\ldots,
r_{J-1} \in D$, with $r_j = j \Delta r$ for $j = 0, \ldots, J-1$. Further, we
approximate the integral using the standard composite
trapezium rule for the quadrature with nodes $\{r_j\}$  and weights $\{ b_j \}$ given
by $b_0 = b_{J-1} = \Dx /2$, and $b_j = \Dx$ otherwise.
Let $U(t)\in\RSet^J$ be the vector approximating
$\big(u(r_0,t),\ldots,u(r_{J-1},t)\big)^T$, $W \in \RSet^{J \times J}$ the matrix
with components $W_{ij}=w(r_i,r_j)b_j$, $S:\R^J\mapsto\R^J$ is the multidimensional
analogous of the firing rate $\sigma$, $S(U(t))$ and $I(t;\theta,a)$ are the vectors
in $\RSet^J$, such that $S(U(t))$ approximates
$\big(\sigma(u(r_0,t)),\ldots,\sigma(u(r_{J-1}))\big)^T$ and
$I(t;\theta,a)=\big(\Xi(r_0,t;\theta,a),\ldots,\Xi(r_{J-1},t;\theta,a)\big)^T$.
This leads to the semi-discrete model
\begin{equation}\label{eq:semi_dis}
    \tau \frac{dU(t)}{dt}=-U(t)+WS(U(t))+I(t; \theta,a), \qquad U(0) = U_0 \in \R^J,
\end{equation}

\begin{table}
  \small
\begin{tabular}{l l c } 
 \hline
 \textbf{Parameter} & \textbf{Description} & \textbf{Value(unit)}   \\ [0.5ex] 
 \hline
$\tau$ & membrane time constant of neurons & $0.03~(\mathrm{s})$\\
 $\ell_{ex}$& spatial spread of of excitatory connections (short-range)&$0.5~(\mathrm{mm})$\\
 $\ell_{in}$&spatial spread of of inhibitory connections (long-range)&$1~(\mathrm{mm})$\\
$A_{ex}$ &excitatory connections strength&$10~(\mathrm{mV})$\\
$A_{in}$& inhibitory connections strength&$6~(\mathrm{mV})$\\
  $\lambda$ & steepness of the sigmoidal, controls firing rate sensitivity &$5~(\mathrm{mV})^{-1}$\\
$\mu$& activity threshold for neural firing&$0.5~(\mathrm{mV})$\\ [1ex] 
 \hline
\end{tabular}
\caption{Model parameters used throughout this paper unless stated otherwise.}
\label{table:params}
\end{table}

As common in sequential data assimilation, we work with a discrete-time approximation
to the dynamical system of \Cref{eq:semi_dis}. Here we integrate \Cref{eq:semi_dis}
in time using a standard explicit Euler method on a uniform grid of the temporal
domain $[0,T]$, hence for a given $K\in\N$ we let $t_k=k\Dt$ with $k=0,1,\ldots,K$
and $\Dt=T/K$, and consider the approximating sequence $U_k \approx U(t_k)$ given by
%
\begin{equation}\label{eq:NDF}
   U_{k+1}=\Psi_k(U_k, \theta) := U_k+\frac{\Delta t}{\tau}\big(-U_k+WS(U_k)+I(t_k;
   \theta,a)\big), \quad U_0 = U(0),
\end{equation}
which we take as our model of cortical dynamics. Note that $\Psi_k$ also depends on other non-inferred parameters such as $a$ and the parameters appearing in the functions $w$ and $\sigma$. For clarity of notation, we omit this dependence when no ambiguity arises.

\subsection{Background on particle filters and Bayesian estimation.}\label{section:filtering} 

\subsubsection{The data assimilation framework}

Data assimilation combines uncertain estimations from a model with noisy (and
sometimes partial) observations to provide an estimate of an unobserved (hidden)
dynamical state. Consider two stochastic processes $\{X_k\}_{k=0}^{{\infty}} \subset
\RSet^{d_x}$ 
and
$\{Y_k\}_{k=1}^{{\infty}} \subset \RSet^{d_y}$,
with realisations $x_k$, $y_k$, which  describe the hidden state and measurements,
respectively. We adopt the standard notation 
$x_{k:l}=\left\{x_k,x_{k+1},\ldots,x_{l}\right\}$ and similarly for $y_{k:l}$. At time step $k\geq 0$, $X_k \in \R^{d_x}$ represents the hidden state to be estimated and for $k\geq 1$, $Y_k \in \R^{d_y}$ represents the measurements. They are assumed to satisfy 
\begin{align}\label{eq:evolution}
  X_{k}&=F_{k}(X_{k-1},\theta,\eta_{k}), & k \geq 1 \\  
        Y_{k}&=G_{k}(X_k,\gamma_k), & k \geq 1.
        \label{eq:observation}  
    \end{align}

The \textit{state equation} \eqref{eq:evolution} describes how the hidden state $X_k$
evolves over time  and the \textit{observation equation} \eqref{eq:observation}
links the hidden state to the measurement $Y_k$.
The model function $F_k:\R^{d_x} \times \RSet^{d_{\theta}} \times \R^{d_x}\to\R^{d_x}$ and observation function $G_k:\R^{d_x}\times \R^{d_y}\to\R^{d_y}$ are assumed to be known. The process noise, $\eta_k \in \R^{d_x}$, represents the model uncertainty and is assumed to have a known probability distribution. Similarly $\gamma_k\in\R^{d_y}$
is the observation noise,  also with a known distribution. We assume that for $k
\neq l$, the process (observation) noise vectors $\eta_k$, $\eta_{l}$
($\gamma_k$, $\gamma_{l}$) are mutually independent and are mutually independent of
the initial state $X_0$. Further, for all $k,l$, $\eta_k$ and $\gamma_{l}$ are
mutually independent. \Cref{eq:evolution,eq:observation} form a \textit{state-space}
system. While the function $F_k$ depends on the control parameters $\theta$, we may
omit this dependence wherever possible from now on.

In this paper, we assume additive noise in the state and observation equations,
and use a linear observation operator for the state variables of \Cref{eq:NDF},
that is, we consider
\begin{equation}\label{eq:SSM_NF}
\begin{aligned}
  X_k&=\Psi_{k-1}(X_{k-1},\theta)+Q\eta_{k} 
          \\  
    Y_k&=G X_k+R\gamma_k,
\end{aligned}
\end{equation}
for $ k = 1, \ldots, K$, where $d_x=d_y=J$, $G\in\R^{J\times J}$ is an observation matrix that is specified later for
each experiment, $\gamma_k,\eta_{k}\sim\mathcal{N}(0,\Id_J)$ are independent and identically distributed normal random variables with values in $\R^J$, and $Q,R \in
\R^{+}$  are the noise intensities.

With a slight abuse of notation we use $\pi(x)$ and $\pi(y)$ to denote the probability
densities of $X$ and $Y$ respectively, even though they are different functions (the argument is used to distinguish them) and assume that the probability distributions of $X_k$ and $Y_k$ are absolutely
continuous with respect to the Lebesgue measure ($\mu_{X_k}(dx_k)=\pi(x_k) dx_k$ and
similarly for $Y_k$).

It is standard to assume $\{X_k\}_{k=0}^{{\infty}}$ is a Markovian process
    \begin{equation}    \pi(x_{k+1}|x_{0:k})=\pi(x_{k+1}|x_k),
    \end{equation}
    and depends on the past observations $y_{1:k}$ only through its own history
    \begin{equation}
        \pi(x_{k+1}|y_{1:k})=\pi(x_{k+1}|x_k).
    \end{equation}

Similarly $\{Y_k\}_{k=1}^{{\infty}}$ is assumed to be a Markovian process with respect to the history of $\{X_k\}_{k=0}^{{\infty}}$, such that
    \begin{equation}
      \pi(y_{k}|x_{1:k})=\pi(y_{k}|x_k). 
    \end{equation}

\subsubsection{Bayesian filtering}

In the Bayesian setting, the sequential filtering problem is formulated as determining
$\pi(x_k|y_{1:k})$ the posterior probability density of the state $X_k$ at time step
$k$. Bayesian filtering methods operate sequentially by incorporating each new
observation using Bayes' formula. At timestep $k+1$, the posterior density of the
state from the previous timestep $\pi(x_k|y_{1:k})$ is evolved in time through the
state equation (\ref{eq:evolution}), leading to the prediction density
$\pi(x_{k+1}|y_{1:k})$, so

\begin{equation}\label{eq:BayesPred}
  \pi(x_{k+1}|y_{1:k})=\int \pi(x_{k+1}|x_k)\pi(x_{k}|y_{1:k})dx_k.
\end{equation}

This prediction is then taken as a prior in the Bayes' formula and is updated or
corrected with the new observation $y_{k+1}$ using the likelihood 
$\pi(y_{k+1}|x_{k+1})$
derived from the observation equation (\ref{eq:observation}), that is
\begin{equation}\label{eq:BayesUpd}
        \pi(x_{k+1}|y_{1:k+1}) =\frac{\pi(y_{k+1}|x_{k+1})\pi(x_{k+1}|y_{1:k})}{\pi(y_{k+1}|y_{1:k})},
    \end{equation}
where 
    \begin{equation*}
        \pi(y_{k+1}|y_{1:k})=\int \pi(y_{k+1}|x_{k+1})\pi(x_{k+1}|y_{1:k})dx_{k+1}.
    \end{equation*}

When the state-space model is linear and Gaussian, the densities in
\eqref{eq:BayesPred} and \eqref{eq:BayesUpd} can be computed exactly leading to the
Kalman filter \cite{kalman1960new}. In the presence of non-linearities, several
approximations exist such as the extended Kalman filter
\cite{Jazwinski1970StochasticPA,gelb1974applied} or the unscented Kalman filter
\cite{UKF}. 
Particle filters \cite{Gordon1993NovelAT,doucet2001sequential} are an alternative approach and are a class of filters that do not require any additional assumptions of linearity or Gaussianity of the state-space model.
They rely only on Monte Carlo methods and importance sampling to approximate the posterior density through an ensemble of random samples, referred to as particles. In practice, particle filters implement the Bayesian filtering \Cref{eq:BayesPred,eq:BayesUpd}: they rely on having an ensemble of $M$ particles $\{x_k^1,\cdots,x_k^M\}$ (positions in the state space at time $k$) along with their weights $\{\omega_k^1,\cdots,\omega_k^M\}$ (the probability that a particle represents the true position) as an approximation to the density $\pi(x_k|y_{1:k})$.
The aim is to first produce an ensemble of particles to approximate the prediction
measure $\pi(x_{k+1}|y_{1:k})$ and then another ensemble for the posterior
$\pi(x_{k+1}|y_{1:k+1})$ with updated weights that reflect how the particles match
the observations. The first step uses the state equation \eqref{eq:evolution} to
approximate \Cref{eq:BayesPred}, while the second makes use of the prior distribution
as an importance sampling distribution in order to make the update formula of the
weights in \Cref{eq:BayesUpd} sequential. A direct implementation will run into a
problem of degeneracy: as time progresses, only few particles will have significant
weight while the  majority of the particles end up with extremely low weights. This
implies that only a few particles contribute significantly to the approximation of
the posterior distribution, effectively reducing the ensemble diversity and the
quality of the estimation (since not all regions of the posterior are sampled).
{\color{black} In fact, at time step $k$ the posterior distribution is approximated
  through a weighted empirical measure 
\begin{equation*}
    \pi(x | y_{1:k})\approx \sum_{m=1}^{M} \omega_k^m \delta(x -x_k^m),
\end{equation*}
where $\delta$ is the Dirac delta. From which we have the estimate 
\[
  \E[g(x)|y_{1:k}]\approx \sum_{m=1}^{M} \omega_k^m g(x_k^m)
\]
,  for any integrable
function of interest $g$ (e.g. the posterior mean or variance). The quality of such estimates
relies on the particles covering the relevant regions of the posterior and on their
weights corresponding to their likelihood. This fails when the ensemble is composed of
few particles with non-negligible weights. }
Indeed, in the worst case, the ensemble may collapse to a single particle with non-negligible
weight. To counter this, a  resampling step is normally added, so that particles that
do not match the observation (and thus have extremely low weight) are thrown away and
replaced with duplicates of the particles with higher probability. In practice,  we
resample our ensemble of particles each time the effective number of particles $\Meff$, approximated by $\Meff:=1/\sum_{m=1}^{M} (\omega_k^{m})^2$, falls
below a threshold of $M/2$. 
{\color{black} $\Meff$ can be interpreted as the number of independent, equally-weighted samples 
that would give the same quality approximation. A high $\Meff$ indicates a diverse particle ensemble, 
for example, when all the particles are uniformly weighted with  $1/M$, $\Meff=M$. On the other hand, when
all weights are zero except for a single particle with weight one, $\Meff=1$ indicating that 
the entire posterior is represented by a single particle, regardless of how many particles were nominally propagated.}
The resampling step consists in drawing the
particles with replacement according to their normalized weight. Based on the findings of \cite{Resamp,sampleRev}, we use stratified resampling in our numerical simulation
as given in \cite{kitagawa1996monte}. 

\Cref{alg:sir} presents a straightforward
implementation of this instance of particle filters, known as the sequential importance
resampling (SIR) particle filter.

\begin{algorithm}
\caption{SIR}
\label{alg:sir}
\begin{algorithmic}[1]
  \REQUIRE Number of particles $M$, state‐space model, observations $\{y_{1:K}\}$, prior $\pi(x_0)$
  \ENSURE Approximate posterior distribution $\pi(x_k\mid y_{1:k})$ at time step $k=1,\ldots,K$ via ensembles of particles $\{x_k^1,\ldots,x_k^M\}$ and their weights $\{\omega_k^1,\ldots,\omega_k^M\}$
  \STATE \textbf{Initialize:} $\{x_0^m\}_{m=1}^M \sim \pi(x_0)$, set weight $\omega_0^m \gets 1/M \;\forall m$
  \FOR{$k=1$ to $K$}
    
    \FOR{$m=1$ to $M$}
      \STATE \label{line:prediction} Sample   $\tilde x_k^m \;\sim\; \pi\bigl(x_k \mid x_{k-1}^m\bigr)$

    
      \STATE Compute importance weight $\tilde \omega_k^m \;\gets\; \omega_{k-1}^m\;\pi\bigl(y_k \mid \tilde x_k^m\bigr)$
      
    \STATE \label{line:norm_weight} Normalize weights 
      $\omega_k^m \;\gets\;\tilde \omega_k^m/\sum_{m=1}^M \tilde \omega_k^m\;$
    \ENDFOR
    \IF{$\Meff=1/\sum_{m=1}^M (\omega_k^m)^2 \leq M/2$}
        \FOR{$m=1$ to $M$}
          \STATE Draw index $l$ from the set $\{1,\dots,M\}$ with $\Pr(l=i)=\omega_k^i$
          \STATE Set  $x_k^m \;\gets\; \tilde x_k^{l},\quad
            \omega_k^m \;\gets\; 1/M$
          
        \ENDFOR
    \ELSE
    \STATE \label{line:end} Set $x_k^m \;\gets\; \tilde x_k^m\;\quad\forall m$
    
    \ENDIF
  \ENDFOR
\end{algorithmic}
\end{algorithm}

\subsubsection{Nested particle filters}

\par In the previous sections, it was implicitly assumed that the model function
$F_k$ is fully known and depends only on the state itself. However, in most
applications, the exact values of the model parameters $\theta$ are not known and
need to be estimated along with the state.

One approach to perform a joint state and parameter estimation is the nested particle filter presented in \cite{Crisan2013NestedPF,Miguez23}, which we adopt here. This nested filter is a recursive algorithm that can be described as two layers of particle filters. In the outer layer, $N$ particles for the parameters are generated, and for each particle, an inner  filter from \Cref{alg:sir} is used to estimate the state with $M$ particles for each of the $N$ parameter particles. 
For the sake of completeness, we present the method in  \Cref{alg:NPF}.

The jittering or rejuvenation step of the algorithm consists in drawing new parameter samples from a kernel to avoid the collapse of the ensemble of particles for the parameters. The new sample $\overline{\theta}_k^i$ must be close enough to the previous sample $\theta_{k-1}^i$, so that the filter approximation of $x_{k-1}$ computed for $\theta_{k-1}^i$ can be used as a particle approximation of the filter for the new sample $\overline{\theta}_k^i$. Following \cite{Crisan2013NestedPF,Miguez23}, we use the following kernel
\begin{equation}\label{eq:JitKer}
    \kappa_k^N(\theta)=(1-\epsilon_N)\delta_{\theta}+\mathcal{TN}(\theta,{\sigma^2}_{\theta}^N, A_{\theta}, B_{\theta}),
\end{equation}
where $\epsilon_N=\frac{1}{\sqrt{N}}$ is the probability of jittering a particle, $\mathcal{TN}$ is the truncated normal distribution with mean $\theta$, variance ${\sigma^2}_{\theta}^N \propto N^{-3/2}$ 
and a support $[A_{\theta}, B_{\theta}]$ that varies for each component of the parameter vector $\theta$.

We consider the empirical posterior means 
$$\E[\theta_k|y_k]\approx\frac{1}{N}\sum_{n=1}^N \theta_k^n, \qquad \E[x_k|y_k]\approx\frac{1}{NM}\sum_{n=1}^N\sum_{m=1}^M x_k^{(n,m)}$$ as our estimates at time step $k$ of the parameters and state respectively.

\begin{algorithm}
\caption{Nested Particle Filter}
\label{alg:NPF}
\begin{algorithmic}[1]
  \REQUIRE 
    Number of parameter particles $N$, number of state particles $M$, parameter prior $\pi(\theta_0)$, state prior $\pi(x_0|\theta_0)$, state-space model, observations $\{y_{1:K}\}$
    \vspace{0.8ex}
  \ENSURE 
    Posterior parameter ensembles of particles and their weights $\{\theta_k^n,v_k^n\}_{n=1}^{N}$ and associated state ensembles $\{x_k^{(n,m)}\}_{n,m}$
  \STATE \textbf{Initialize:}
  \FOR{$n=1$ to $N$}
    \STATE Sample $\theta_0^n\sim \pi(\theta_0)$, set $v_0^n\gets 1/N$
    \vspace{0.8ex}
      \STATE Sample $\{x_0^{(n,m)}\}_{m=1}^M\sim \pi(x_0\mid\theta_0^n)$, set $\omega_0^{(n,m)}\gets 1/M \;\forall m$
  \ENDFOR

  \FOR{$k=1$ to $K$}
    \FOR{$n=1$ to $N$}
    \STATE \textbf{Jitter}
    sample $\overline{\theta}_k^n \;\sim\; \kappa_k^N(\theta_{k-1}^n)$

      \FOR{$m=1$ to $M$}
        \STATE Sample $x_k^{(n,m)} \;\sim\; \pi\bigl(x_k\mid x_{k-1}^{(n,m)},\,\overline{\theta}_k^n\bigr)$
        
        \STATE Compute unnormalized state weights $\tilde \omega_k^{(n,m)} \;\gets\; \omega_{k-1}^{(n,m)}\;\pi\bigl(y_k\mid x_k^{(n,m)},\,\overline{\theta}_k^n\bigr)$
        
      \ENDFOR
      \STATE Estimate parameter likelihood $\hat L_k^n \;\gets\; \sum_{m=1}^{M}\tilde \omega_k^{(n,m)}$
      \vspace{0.5ex}
      
      \STATE Run SIR steps \ref{line:norm_weight} to \ref{line:end} on the ensemble $\{x_k^{(n,m)},\tilde \omega_k^{(n,m)}\}_{m=1}^{M}$
      \vspace{0.5ex}
      
      \STATE Update unnormalized parameter weight $\tilde v_k^n \;\gets\; v_{k-1}^n\;\hat L_k^n$
      
    \ENDFOR
    \STATE Run SIR steps \ref{line:norm_weight} to \ref{line:end} on the ensemble $\{\overline{\theta}_k^n,\tilde v_k^n\}_{n=1}^{N}$

  \ENDFOR
\end{algorithmic}
\end{algorithm}

\section{Nested particle filter applied to synthetic data }\label{section:synthetic}

Before using the particle filter on cortical data, we test the data-assimilation
algorithm on synthetic data. We consider the states $\{ U^*_k \}_{k = 0}^{K^*}
\subset \RSet^{J^*}$ generated by model \cref{eq:NDF}, in which the mapping
$\Psi^*_k(U,\theta^*)$ is specified via a matrix $W^* \in \RSet^{J^*
\times J^*}$ and input $I^*(t; \theta^*, a^*) \in \RSet^{J^*}$.

The input features 3 inferable parameters $\theta^* = (A^*, \nu^*, f^*)$ as well as an
additional parameter $a^*$, and has components $\left(\Xi(r_j,t,\theta^*,a^*)\right)_{j=0}^{J^*-1}$ at time $t$ and spatial nodes $r_j=j\Dx$ where
\[ 
  \Xi(r,t,\theta^*,a^*) = A^* \cos 2 \pi \bigl(\nu^* r - (f^* + a^* t) t\bigr).
\]
The external input models a \textit{chirp travelling wave} with amplitude $A^*$ (mV),
spatial frequency $\nu^*$ ($\textrm{mm}^{-1}$), and time-dependent frequency $f(t) =
(f^* + a^*t)$, hence $f^*$ (Hz) is the initial temporal frequency and  $a^*$ (Hz/s) is the chirp
rate.

The data-assimilation task aims to reconstruct $\{  U^*_k \}$ and infer $\theta^*$
under noisy observations, as well as (epistemic) model uncertainty. To this
end, we setup system  \cref{eq:SSM_NF} as follows:
\begin{itemize}
  \item \textit{Noisy observations.} The observations $\{ Y_k \}_{k=1}^K
    \subset \RSet^J$ are obtained by sampling the original signal $\{ U^*_k
    \}_{k=0}^{K^*} \subset \RSet^{J^*}$ on a coarse spatio-temporal grid ($K < K^*$,
    $J < J^*$), and by adding noise.
  \item \textit{Model uncertainty.} When performing data assimilation, one has
    typically only limited knowledge of the model that generated the data.
    To mimic this scenario, we use a mapping $\Psi_k(\cdot, \theta)$ in the state space system \cref{eq:SSM_NF} that differs from $\Psi_k^*(\cdot,\theta^*)$ used to generate the original state in two ways:
    firstly, the matrix $W\in \RSet^{J \times J}$ and input vector 
    $I(t;\theta,a) \in \RSet^J$ sample the functions $w$ and $\Xi$ on the coarse
    spatial grid; secondly, the input vector $I(t;\theta,a)$ has components $\bigl(\Xi(r_j,t,\theta,0)\bigr)_{j=0}^{J-1}$, that is, the model used for inference has null chirp rate
    and constant frequency $f$, while the one giving rise to the observations has a
    time-dependent frequency. {\color{black} Although it is beyond the scope of this work and not necessary for the comprehension and interpretation of the results presented here, we provide in \Cref{appendix: sensitivity} a minimal quantification of the relative influence of the inferable parameters $\theta$ through first and total order Sobol indices.}
\end{itemize}

  We present numerical evidence that the nested particle filter algorithm generates
  estimating sequences of states $\{ X_k \}$ and empirical posterior means of parameters $\{ \theta_k \}$ such
  that: the state $\{ X_k\}$ is close
  to $\{ U^*_k \}$ on the coarse spatio-temporal grid; the sequence $\{ (A_k,\nu_k)
  \}$ converges to  $(A^*,\nu^*)$; the sequence $\{ f_k \}$ does
  not converge to a limit, but nevertheless provides some information on the original
  model. Our numerical experiment is setup as follows.
\begin{itemize}
  \item \textit{Generating the signal.}
    To generate $\{ U^*_k \}$ we simulate \cref{eq:NDF} on $(r,t) \in [0,L] \times [0,T]$ with $L =
    10\textrm{mm}$ and $T = 100s$, using a grid with $J^* = 500$ spatial
    points, and $K^* = 20000$ time points. The initial condition is given by $u_0(r)
    = \sin(\pi r)$, and model parameters are fixed as
    \[
      (\theta^*,a^*) = (A^*,\nu^*,f^*,a^*)=(1,0.1,0.5,-0.005). 
    \]
  \item \textit{Generating observations.}
    We interpolate $\{ U^*_k \}$ on a coarse grid of $[0,L]\times[0,T]$ with $J = 30$ and $K= 10000$, and obtain an
    intermediate sequence $\{ \bar U_k \}_{k =0}^K \subset \RSet^{J}$. The
    observation vectors satisfy
    \[
      Y_k = \bar U_{k} + \epsi \xi_k, 
      \qquad \xi_k \stackrel{i.i.d}{\sim} \mathcal{N}(0,\Id_J), 
      \qquad k = 1,\ldots,K,
      \qquad \epsi = 0.5.
    \]
  \item \textit{Running the nested particle filter.} In the state-space system
    \cref{eq:SSM_NF} we set
    \[
    Q = 0.1, 
    \qquad 
    R = 0.5, 
    \qquad \eta_k, \gamma_k \stackrel{i.i.d}{\sim} \mathcal{N}(0,\Id_J), 
    \qquad k = 1,\ldots,K.
    \]
    Further, we run \Cref{alg:NPF} once using $N=M=500$ particles with uniform priors on
    the inferable parameters given by: 
    \begin{equation*}
        \theta = (A,\nu,f)\sim \U([0,10])\otimes\U([0,1])\otimes\U([0,1]).
    \end{equation*}
    Priors on the state $X$ are samples from a discretized mean-zero Gaussian
    process with a squared exponential covariance $\mathcal{GP}\big(0,k(r,s)\big),$
    where $k(r,s)=2\exp{\left(-(r-s)^2/2\right)}$, hence we set
     \begin{equation}\label{eq:GP}
         X_0\sim \mathcal{N}(0,C), \qquad C\in\R^{J\times J}, \qquad
         C_{i,j}=k(r_i,r_j).
     \end{equation}
\end{itemize}

\begin{figure}
   
    \begin{subfigure}[b]{.324\linewidth} \includegraphics[width=\linewidth]{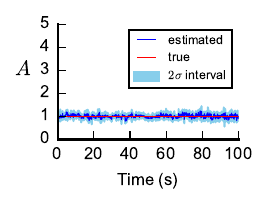}
    \subcaption{}
    \label{fig:syn_I0}
    \end{subfigure}
    \begin{subfigure}[b]{.324\linewidth} \includegraphics[width=\linewidth]{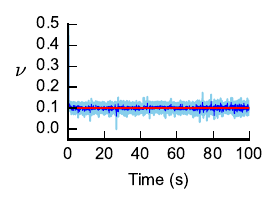}
    \subcaption{}
    \label{fig:syn_j0}
    \end{subfigure}
    \begin{subfigure}[b]{.324\linewidth} \includegraphics[width=\linewidth]{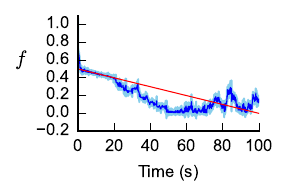}
    \subcaption{}
    \label{fig:syn_f0}
    \end{subfigure}
    \vfill
    \begin{subfigure}[b]{.324\linewidth} \includegraphics[width=\linewidth]{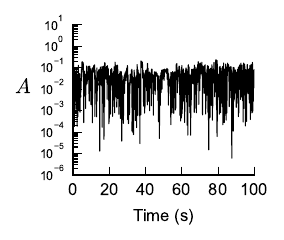}
    \subcaption{}
    \label{fig:syn_I0_err}
    \end{subfigure}
    \begin{subfigure}[b]{.324\linewidth} \includegraphics[width=\linewidth]{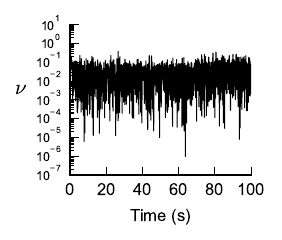}
    \subcaption{}
    \label{fig:syn_j0_err}
    \end{subfigure}
    \begin{subfigure}[b]{.324\linewidth} \includegraphics[width=\linewidth]{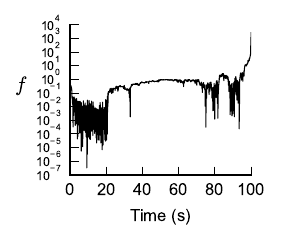}
    \subcaption{}
    \label{fig:syn_f0_err}
    \end{subfigure}
    
    \caption{Parameter estimation of the nested particle filter for the data
      assimilation task with synthetic data, as described in \cref{section:synthetic}.
      (a): The sequence of empirical posterior means $\{A_k\}_k$ (blue)
      {\color{black} with $\pm 2\sigma$ uncertainty band (shaded region)}
      is compared to the true value $A^*$ (red),
      showing that the parameter is inferred {\color{black} almost instantly after a
      very brief burn-in period.}
      (b): Similar to (a), but for the estimation of the parameter $\nu^*$. (c): In the
      model adopted for inference in \cref{eq:SSM_NF}, the
      frequency $f$ is a parameter, while in the one used for generating the data the
      frequency is time-dependent ($f^* + a t_k$). The estimated parameter $f_k$ is
      not constant, as expected, but follows the trend of the time-dependent true
      variable. (d--f): The relative error for each quantity is plotted as a function
    of time in a log-linear scale.}
    \label{fig:syn_cos_params}

\end{figure}
\begin{figure}
    \begin{subfigure}[b]{.49\linewidth} \includegraphics[width=\linewidth]{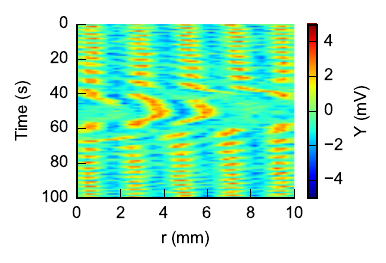}
    \subcaption{Observations}
    \label{fig:obs}
    \end{subfigure}
    \begin{subfigure}[b]{.49\linewidth} \includegraphics[width=\linewidth]{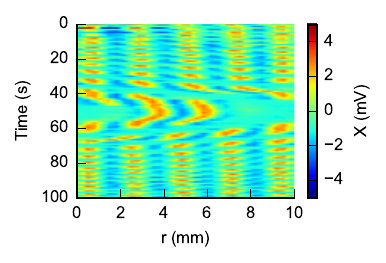}
    \subcaption{Estimated state}
    \label{fig:state}
    \end{subfigure}
    \vfill
    \begin{subfigure}[b]{\linewidth}
    \includegraphics[width=\linewidth]{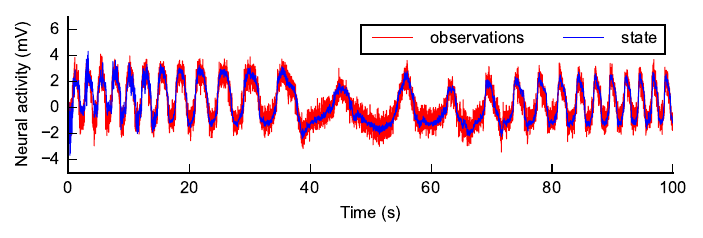}
    \subcaption{$r=5\mathrm{mm}$}\label{fig:time_syn_cos}
    \end{subfigure}
    
    \caption{For the same numerical experiment as \cref{fig:syn_cos_params}, we show
      the observed spatio-temporal data $\{ Y_k \}$ in (a), and the estimated state
      variable $\{ X_k \}$ in (b), displaying a good agreement across all
      spatio-temporal domain. (c): Time traces at point $r=5\mathrm{mm}$ in the
    cortex, as a function of time.}
    \label{fig:syn_state_obs}
\end{figure}

\Cref{fig:syn_I0,fig:syn_j0,fig:syn_f0} show the sequence of empirical posterior means of parameters $\{(A_k, \nu_k, f_k)\}$ {\color{black} and $\pm 2\sigma$ uncertainty band} in relation to $\{ (A^*, \nu^*, f^* + a^* t_k)\}$, while in
\cref{fig:syn_I0_err,fig:syn_j0_err,fig:syn_f0_err} we examine the time
evolution of the relative error in each component. In \cref{fig:syn_I0,fig:syn_j0}
{\color{black} we observe that the posterior means for the
amplitude $A$ and spatial frequency $\nu$ converge almost immediately after a very
brief burn-in period, with a small and stable uncertainty band.}
This is confirmed by
\cref{fig:syn_I0_err,fig:syn_j0_err} where we
see relative errors of the order of $10^{-1}$ for these two parameters. As for the
temporal frequency $f$, we observe in
\Cref{fig:syn_f0} that 
{\color{black} during the first $20\mathrm{s}$, the posterior mean
  tracks closely the true dynamics, then it progressively diverges below
the truth between $20\mathrm{s}$ and $50\mathrm{s}$, after this, the estimate increases gradually and stays relatively close to the truth. The $2\sigma$ bans remains small and stable during the whole observation window.}

The particle filter thus estimates correctly parameters that can be meaningfully
mapped from the original model $\psi^*$ to the one used for inference, $\psi$. A user
with knowledge of $ \{ Y_k \}$ and $\Psi^*$, may interpret the  lack
of convergence in $\{f_k\}$ as model misfit, while in reality this can be used to refine $\Psi$ used in inference. In fact, the model $\Psi^*$ which gave rise to
the observations has a time-dependent
frequency $f(t)$ whose law is not explicitly included in the model $\Psi$, and the sequence $\{
f_k \}$ tries to approximate it for some time. Through propagation, weighting and resampling of the parameter particles, the filter manages to align the estimated parameter to the observed behaviour of the state dynamics. This can be helpful in refining the state-space model to better align with the observations, by incorporating an additional dynamical component for $f$ in $\Psi$, suggested by the evolution of the sequence $\{f_k\}$.

In \Cref{fig:syn_state_obs} we show the observations (\Cref{fig:obs}) and estimated
state (\Cref{fig:state}) along with the time evolution in \Cref{fig:time_syn_cos} of
both the observations and estimated state at $r=5\mathrm{mm}$. In the time series of
\Cref{fig:time_syn_cos}, the falling then rising of the temporal frequency of the state,  {\color{black} as well as the slowing of the dynamics between $40\mathrm{s}$ and $60\mathrm{s}$}, are plainly
visible (see also \Cref{fig:syn_f0}). 

To assess the state-estimation error over the
spatio-temporal grid, we consider the aggregated
root mean square error (RMSE) 
\begin{equation}\label{eq:rmse}
    \rmse=\sqrt{\frac{1}{KJ}\sum_{k=1}^{K}\|Y_k-X_k\|^2}.
      \end{equation}
For this synthetic data experiment, we found $\rmse=0.561\mathrm{mV}$.

\begin{figure}
    \begin{subfigure}[b]{.648\linewidth} \includegraphics[width=\linewidth]{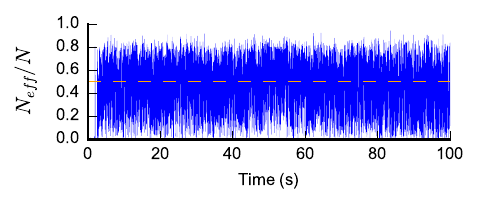}
    \subcaption{}
    \label{fig:ess}
    \end{subfigure}
    \begin{subfigure}[b]{.324\linewidth} \includegraphics[width=\linewidth]{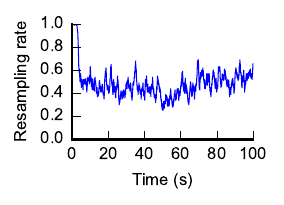}
    \subcaption{}
    \label{fig:resampling_rate}
    \end{subfigure}
    \caption{For the same numerical experiment as \cref{fig:syn_cos_params} we show
    the sequence of the ratio $N_\textrm{eff}/N$ in (a) and the rolling resampling rate over a window of 100 time steps in (b).}
    \label{fig:wrong_prior_params}
\end{figure}

{\color{black}
To evaluate particle diversity and identify potential degeneracy, we
visualize in \cref{fig:ess} the ratio $N_{\textrm{eff}}/N$ which indicates how many
independent samples
 the parameter particle population is effectively equivalent to. A ratio close to one indicates 
a well-spread sample, while a value close to zero indicates weight degeneracy and
happens when the ensemble collapses.
The dashed line indicates the resampling threshold of $N_{\textrm{eff}}<N/2$. The
ratio is near zero 
in the burn-in period, which may indicate that the initial particle distribution does not match 
the true parameters. These early estimates are not reliable and should be discarded.
After the burn-in period, the
filter recovers and oscillates rapidly in the range $0-0.9$ with no trend, which indicates frequent resampling. 
In addition to its role in maintaining sample diversity, this frequent resampling, together with jittering, is a mechanism via which the ensemble of parameter particles moves 
around in order to account for the dynamics of the true parameters (the frequency $f$) that 
were not included in the state-space model. This is further confirmed by \cref{fig:resampling_rate}, which shows the rolling sampling rate over a window of $100$ time steps. At the start, 
    the rate is one which corresponds to the early weight degeneracy discussed before, and indicates resampling at almost each time step during this early stage.
    After that, the resampling rate settles into a stable regime in the range $0.3-0.7$ With a dip in the period $40-60\mathrm{s}$ corresponding to slower dynamics.}

{\color{black} While our approach showed its capability to handle joint  estimation of (potentially non static) parameters and
state of a nonlinear state space model, several limitations need to be pointed out.  Particle filter methods suffer from the curse of dimensionality \cite{ObstaclestoHighDimensionalParticleFiltering,Bickel2008SharpFR}, where the number of particles needed to accurately approximate 
the posterior distribution grows rapidly with the dimension of the state, often leading to particle degeneracy and increased computational cost. 
This holds to an even greater extent for the nested particle filter of \Cref{alg:NPF} , which relies on two layers of filters. 
Although a 30-dimensional system of the experiment above is not prohibitively large, increasing the state dimension causes rapid growth in particle requirements and computational cost.
Additionally, performance is highly sensitive to design choices, most notably the noise intensities $Q$ and $R$, the resampling scheme and the jittering kernel (\cite{Arulampalam_tutorial,Doucet_tutorial}). 
Misspecified $Q$ and $R$ can result in a collapsed or excessively diffuse ensemble, poor resampling settings can reduce diversity or waste computation, 
while a jittering kernel with poorly calibrated variance and support can hinder the convergence to the parameter posterior. We also note that 
discretisation choices when defining the state-space model can introduce further design dependent error.
Based on this, the application of such methods requires careful problem-specific tuning to achieve reliable performance. In our case, the noise intensities were chosen
by running the algorithm repeatedly over a set of candidate values and  visually inspecting the spread of the ensemble of particles and the inferred parameters trajectories; the pair yielding the lowest RMSE while maintaining a high effective sample size was chosen. 
The discretisation steps were required to allow us to resolve the frequencies present in the observations, and the resampling scheme was chosen based 
on the results reported in the literature. For additional experiments highlighting the effect of some of these design choices, see \Cref{Appendix:discrete,Appendix:noise,Appendix:prior}.}

\section{Experimental Cortical Data}\label{section:data} 
We now make some steps towards performing data assimilation on a recording of the
local field potential (LFP) from the cortex of a mouse during natural
sleep \cite{pedrosaHippocampalGammaSharp2022}, {\color{black} which we take as observations of coarse cortical activity
corrupted by noise}. In this section we describe preliminary data analysis that was
performed prior to data assimilation. We denote the LFP measurements by $V$.

The data provided were labelled into three
categories: rapid eye movement (REM) sleep, slow wave sleep (SWS) and unclassified,
corresponding to non-specified resting states. The recording has a duration of
$T=900\mathrm{s}$ with an acquisition rate of $50\mathrm{Hz}$, yielding $45000$ observations with $\delta t=0.02\mathrm{s}$.

\begin{figure}[h!]
    \centering
            
    \begin{subfigure}[b]{.324\textwidth}
    
\includegraphics[width=\linewidth]{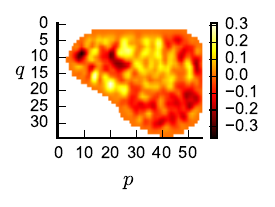}

        \subcaption{}
        \label{fig:t200_2d}
    \end{subfigure}
    \begin{subfigure}[b]{.324\textwidth}   \includegraphics[width=\linewidth]{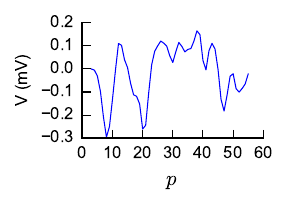}
        \subcaption{}
        \label{fig:t200_j10}
    \end{subfigure}
    \begin{subfigure}[b]{.324\textwidth}   \includegraphics[width=\linewidth]{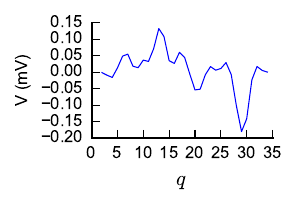}
        \subcaption{}
        \label{fig:t200_i40}
    \end{subfigure}

    \vfill
    
    \begin{subfigure}[b]{.324\textwidth}
\includegraphics[width=\linewidth]{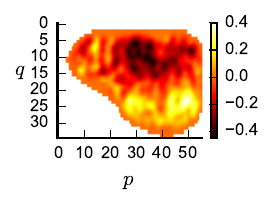}
        \subcaption{}
        \label{fig:t600_2d}
    \end{subfigure}
    \begin{subfigure}[b]{.324\textwidth}   \includegraphics[width=\linewidth]{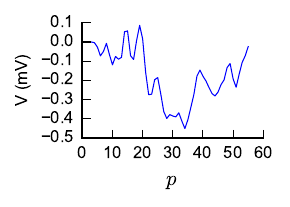}
        \subcaption{}
        \label{fig:t600_j10}
    \end{subfigure}
    \begin{subfigure}[b]{.324\textwidth}   \includegraphics[width=\linewidth]{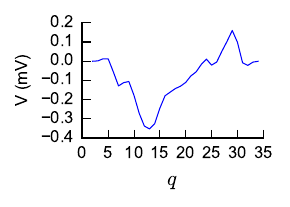}
        \subcaption{}
        \label{fig:t600_i40}
    \end{subfigure}
    
    \vfill
    
    \begin{subfigure}[b]{.324\textwidth}
\includegraphics[width=\linewidth]{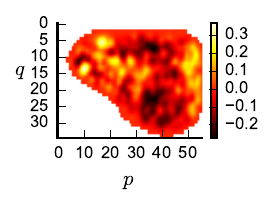}
        \subcaption{}
        \label{fig:t850_2d}
    \end{subfigure}
    \begin{subfigure}[b]{.324\textwidth}   \includegraphics[width=\linewidth]{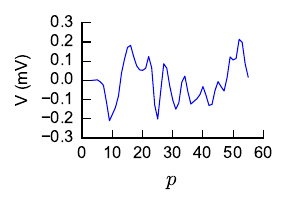}
        \subcaption{}
        \label{fig:t850_j10}
    \end{subfigure}
    \begin{subfigure}[b]{.324\textwidth}   \includegraphics[width=\linewidth]{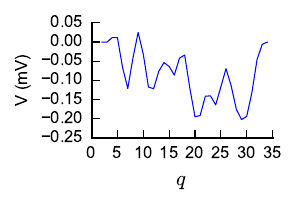}
        \subcaption{}
        \label{fig:t850_i40}
    \end{subfigure}
  \caption{Panels show screenshots from the LFP measurements at three distinct time points corresponding to different sleep stages.
  Each row represents a different time, progressing from earliest at the top to latest at the bottom: (a-b-c) are screenshots taken at $t=200\mathrm{s}$ during an undefined rest stage, (d-e-f) are taken at $t=600\mathrm{s}$ during slow wave sleep (SWS) and (g-h-i) at $t=850\mathrm{s}$ during rapid eye movement sleep (REM). The columns show three representations in space of the measurements, highlighting the global structure as well as two spatial profiles.
  By column: the left column (a-d-g) shows the 2d cortical recordings in the pixel domain $(p,q)$, the middle column (b-e-h) a horizontal slice at pixel index $q=10$ and the right column (c-f-i) a vertical slice at $p=40$. This arrangement enables simultaneous visualization of the 2d field along with slices across two principal directions during different sleep stages.} 
  \label{fig:A4_2d_x}
\end{figure}

In the following, each pixel is indexed by
$(p,q)$, where $p$ and $q$ are its horizontal and vertical placement in the grid respectively.  In \Cref{fig:A4_2d_x} we plot the LFPs from the cortical recording at times $t=200$, $t=600$ and $t=850$ seconds which are data points in the unclassified, SWS and REM sleep stages respectively. 
In \Cref{fig:t200_2d,fig:t600_2d,fig:t850_2d} we plot the recordings on the pixel domain $(p,q)$. We note that at the edge of the recording domain the LFPs is zero. This is further evident in \Cref{fig:t200_j10,fig:t200_i40,fig:t600_j10,fig:t600_i40,fig:t850_j10,fig:t850_i40} which show the LFPs through a vertical (horizontal) slice at $p=40$ ($q=10$).
Next we consider the temporal dynamics and dominant frequencies observed in the data before returning to the spatial dynamics.

\begin{figure}[h!]
    \centering
    \begin{subfigure}[b]{\textwidth}
\includegraphics[width=\linewidth]{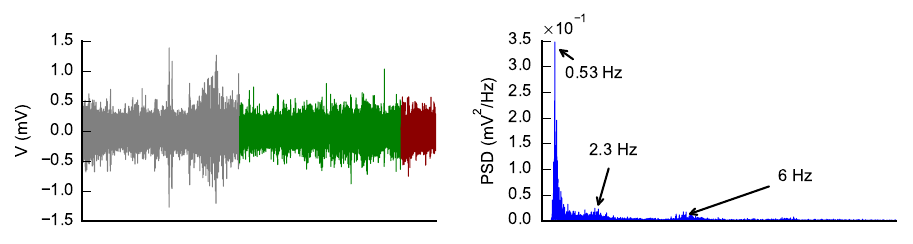}
        \subcaption{$p$=20, $q$=10}
    \end{subfigure}
    \begin{subfigure}[b]{\textwidth}   \includegraphics[width=\linewidth]{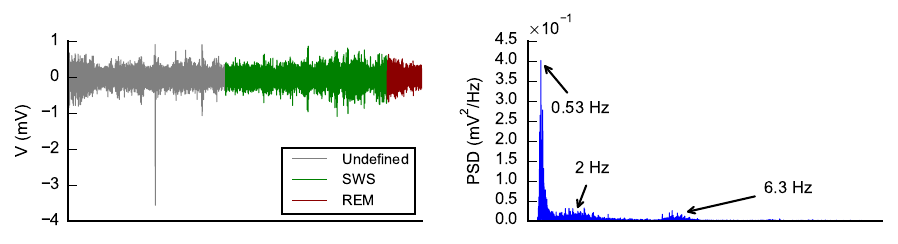}
        \subcaption{$p$=40, $q$=10}
    \end{subfigure}
    \begin{subfigure}[b]{\textwidth}   \includegraphics[width=\linewidth]{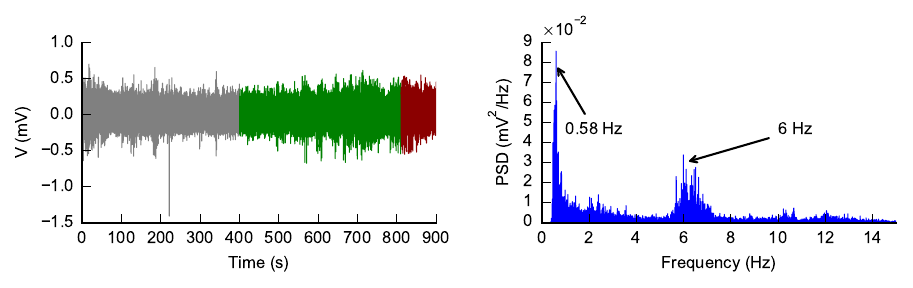}
        \subcaption{$p$=40, $q$=25}
    \end{subfigure}
    \caption{Panels illustrate the temporal dynamics as well as frequency content of
    three pixels in different locations. Each row shows the time series of the neural
  activity of a specific pixel on the left, along with its power spectrum density
(PSD) and highlighted frequency peaks on the right. (a) corresponds to a pixel on the
upper left of the domain $(p=20,q=10)$, (b) corresponds to a pixel on the upper right
$(p=40,q=10)$ and (c) to a pixel on the lower right $(p=40,q=25)$. The three panels
shows similar temporal dynamics between the three pixels, as well as similar
frequency profiles with peaks around $0.5\mathrm{Hz}$, $2\mathrm{Hz}$ and
$6\mathrm{Hz}$.}
    \label{fig:A4_2d_t}
\end{figure}

To illustrate the temporal dynamics we select
pixels at different
locations in the cortex. These are representative of the kind of temporal dynamics present
in the data and are located on the upper left ($p=20,q=10$), upper
right ($p=40,q=10$) and lower right ($p=40,q=25$) of the cortex. The time series, along with their
power spectrum densities (PSD) computed from a single periodogram, are shown in \Cref{fig:A4_2d_t}. We observe a
zero-mean
oscillatory signal that has a dominant frequency component at $0.5\mathrm{Hz}$ and
two broad peaks at around $2\mathrm{Hz}$ and $6\mathrm{Hz}$. The sharp and dominant
peak at $0.5\mathrm{Hz}$ indicates that energy of the signal is concentrated in the
lower frequencies $0-1\mathrm{Hz}$, while the broad peaks around $2\mathrm{Hz}$ and
$6\mathrm{Hz}$ may suggest that the underlying process is frequency-modulated or has
a frequency that fluctuates over time. 
The peak at $6\mathrm{Hz}$ seems to be more prominent for the pixel located on the lower right of the cortex ($p=40,q=25$), this may be explained by the higher curvature of the cortex in this region, which introduces high-frequency artifacts in the recordings.

\begin{figure}[h!]
    \centering
    \begin{subfigure}[b]{.45\textwidth}
\includegraphics[width=\linewidth]{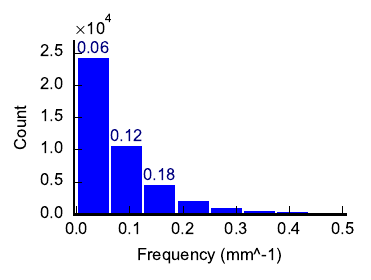}
        \subcaption{$q=10$}
        \label{fig:A4_1d_j10}
    \end{subfigure}
    \begin{subfigure}[b]{.45\textwidth}
\includegraphics[width=\linewidth]{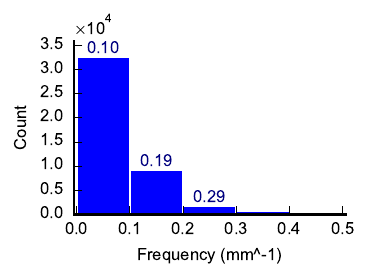}
        \subcaption{$p=40$}
        \label{fig:A4_1d_i40}
    \end{subfigure}   
    
    \caption{Histograms showing the distribution in time of the dominant spatial frequencies present in two slices at $q=10$ (a)  and $p=40$ (b). At each time point, the dominant spatial frequency is extracted by computing the frequency at which the power spectrum density peaks. These peaks are aggregated into a histogram to show their distribution in time. The two histograms exhibit that the temporal evolution for both slices is dominated by larger wavelengths while higher frequency features occur rarely.}
    \label{fig:A4_1d_x_freq}
\end{figure}

Next, we consider the spatial frequencies present across the slices at $q=10$ and
$p=40$. At each time point $t_k$, $k=1,\ldots, 45000$, we extract the dominant
frequency by finding the peak of the PSD in space. \Cref{fig:A4_1d_x_freq}
shows the resulting histograms that aggregate these peaks over the total duration of
the recording. In \Cref{fig:A4_1d_j10} we
examine the horizontal slice at $q=10$ and in \Cref{fig:A4_1d_i40} the vertical slice
at $p=40$. The fundamental frequency for the slice at $q=10$ is
$0.06\mathrm{mm}^{-1}$ and $0.1\mathrm{mm}^{-1}$ for the one at $p=40$. Both histograms show a higher concentration of peak spatial frequencies at lower values. This pattern suggests that large scale spatial features (long wavelengths) dominate, while finer scale features (high frequencies) are transient.

\subsection{Reduction from two to one spatial dimensions}\label{sec:radon}

Since applying the particle filters of \Cref{section:filtering} requires
time-stepping a large number of independent high-dimensional systems of ODEs, we consider reduction from two to one spatial dimension to reduce the computational cost. This enables us to apply the neural mass model of \eqref{eq:semi_dis} from the discretization of the one-dimensional Amari neural field model \eqref{eq:CNF}.
To this end we use the Radon transform \cite{Radon1917} which, as described below, allows us to consider a family of one-dimensional reductions of the
dataset, parametrized by an angle $\alpha$.

To introduce the Radon transform we let $n_{\alpha}=(\cos(\alpha),\sin(\alpha))$ be the
unit vector with angle $\alpha$ from the $x$-axis. For fixed $d\in \R$ and $\alpha \in
[0,\pi)$ we introduce 
\[
    L_{d,\alpha}=\{(r,s)\in \R^2: \langle (r,s), n_{\alpha} \rangle =d\},
\]
that is, the perpendicular to $n_{\alpha}$ whose distance from the origin is $d$.
The Radon transform of a square-integrable function $g \colon \RR \to \RSet$ is the
function $\mathcal{R}g:\R \times [0,\pi) \mapsto \R $ defined as:
\begin{equation}\label{Radon_def}
  \mathcal{R}g:(d,\alpha) \mapsto\int_{L_{d,\alpha}} g(z)dS(z)=\int_{\R}\int_{\R}g(r,s)\delta(\langle (r,s), n_{\alpha}\rangle -d)drds.
\end{equation}

\begin{figure}[h!]
    \centering
            
    \begin{subfigure}[b]{.3\textwidth}
    
\includegraphics[width=\linewidth]{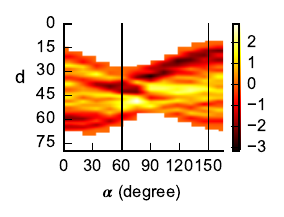}

        \subcaption{}
        \label{fig:t200_sino}
    \end{subfigure}
    \begin{subfigure}[b]{.3\textwidth}   \includegraphics[width=\linewidth]{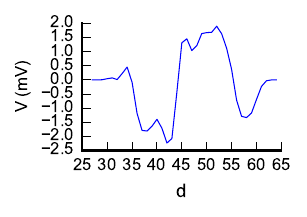}
        \subcaption{}
        \label{fig:t200_angle60}
    \end{subfigure}
    \begin{subfigure}[b]{.3\textwidth}   \includegraphics[width=\linewidth]{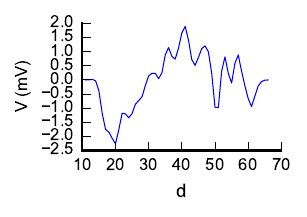}
        \subcaption{}
        \label{fig:t200_angle150}
    \end{subfigure}

    \vfill
    
    \begin{subfigure}[b]{.3\textwidth}
\includegraphics[width=\linewidth]{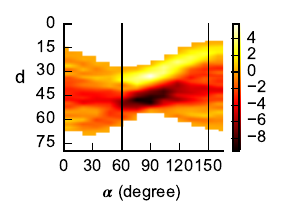}
        \subcaption{}
        \label{fig:t600_sino}
    \end{subfigure}
    \begin{subfigure}[b]{.3\textwidth}   \includegraphics[width=\linewidth]{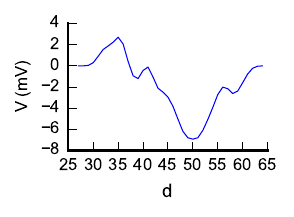}
        \subcaption{}
        \label{fig:t600_angle60}
    \end{subfigure}
    \begin{subfigure}[b]{.3\textwidth}   \includegraphics[width=\linewidth]{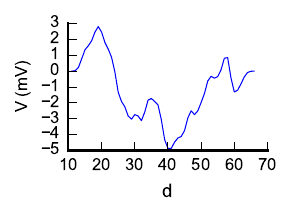}
        \subcaption{}
        \label{fig:t600_angle150}
    \end{subfigure}
    
    \vfill
    
    \begin{subfigure}[b]{.3\textwidth}
\includegraphics[width=\linewidth]{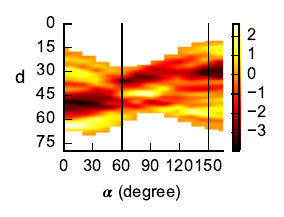}
        \subcaption{}
        \label{fig:t850_sino}
    \end{subfigure}
    \begin{subfigure}[b]{.3\textwidth}   \includegraphics[width=\linewidth]{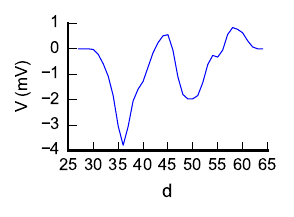}
        \subcaption{}
        \label{fig:t850_angle60}
    \end{subfigure}
    \begin{subfigure}[b]{.3\textwidth}   \includegraphics[width=\linewidth]{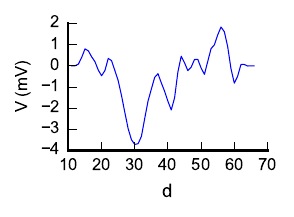}
        \subcaption{}
        \label{fig:t850_angle150}
    \end{subfigure}
  \caption{Panels show screenshots from the Radon transform of the LFP measurements at three distinct time points corresponding to different sleep stages.
  Each row represents a different time, progressing from earliest at the top to latest at the bottom: (a-b-c) are screenshots taken at $t=200\mathrm{s}$ during an undefined rest stage, (d-e-f) are taken at $t=600\mathrm{s}$ during slow wave sleep (SWS) and (g-h-i) at $t=850\mathrm{s}$ during rapid eye movement sleep (REM). The columns show three different representations of the measurements in Radon space.
  The left column (a-d-g) shows the Radon transform of the 2d measurements in the $(d,\alpha)$ plane, referred to as sinogram. The middle column (b-e-h) shows a slice through the sinograms of the left column at $\alpha=60^\circ$, while the right column (c-f-i) corresponds to a slice at $\alpha=150^\circ$. This arrangement enables visualization  during different sleep stages of the full Radon transform along with its spatial profiles for two directions.} 
  \label{fig:A4_radon_x}
\end{figure}

\begin{figure}[h!]
    \centering
    \includegraphics[width=0.5\linewidth]{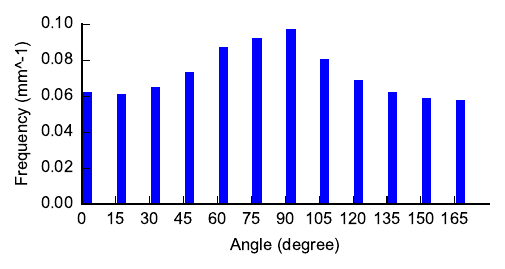}
    \caption{Fundamental spatial frequency in the Radon transformed signal for different angles. Similar to \Cref{fig:A4_1d_x_freq}, we construct histograms of the distributions in time of the dominant spatial frequencies present in slices across each angle. Then the most frequently observed frequency of each histogram is taken as the fundamental frequency for that angle. }
    \label{fig:A4_radon_x_freq_angles}
\end{figure}

That is the collection of integrals of $g$ along the parametrized lines in the plane.
Note that in practice the discretized offset samples in the Radon space are given by
$d=\Dx (p'\sin(\alpha)+q'\cos(\alpha))$, where the indices $p',q'$ are obtained by
centering the pixel domain $(p,q)$ at the origin $(0,0)$. For the remainder of this
work, we take the Radon domain as our one-dimensional cortical domain and fix
$\Dx=0.3\mathrm{mm}$ for all values of $\alpha$. Our choice of this value is
motivated by measurements of the mouse brain reported in the literature 
(for example
\cite{natt_high-resolution_2002}). In fact, this choice results in spatial domains
with lengths ranging from around $9\mathrm{mm}$ ($\alpha=90^{\circ}$) to $16 \mathrm{mm}$
($\alpha=0^{\circ})$, in agreement with the literature.
When $\mathcal{R}g$ is visualized in the $(d,\alpha)$ plane (\Cref{fig:A4_radon_x}),
it typically looks like a superposition of sine and cosine waves and is therefore
called a sinogram \cite{Deans1983TheRT,Toft1996TheRT}. In \Cref{fig:A4_radon_x}
(left), we show the sinogram  at different times (and sleep stages) for the data presented earlier in \Cref{fig:A4_2d_x}. We also plot slices of the sinogram at angles $60^{\circ}$ and $150^{\circ}$ that show the reduced 1D
spatial signal resulting from the Radon transform at these angles. Compared to
\Cref{fig:A4_2d_x} we notice that the 1D signals appear smoother than the original
slices of the 2D signals as sharper features are averaged out. 
We also examine how the fundamental spatial frequency changes with the angle $\alpha$ in \Cref{fig:A4_radon_x_freq_angles}. We observe that it changes only slightly with the projection angle, which suggests that the original 2D signal is mostly isotropic and its underlying spatial frequency content does not depend on orientation.  
 
\begin{figure}[!h]
    \centering
    \begin{subfigure}[b]{\textwidth}
\includegraphics[width=\linewidth]{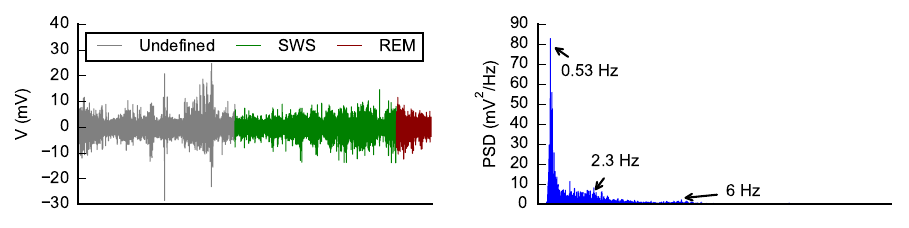}
        \subcaption{$\alpha=60^{\circ}$}
    \end{subfigure}
    \begin{subfigure}[b]{\textwidth}   \includegraphics[width=\linewidth]{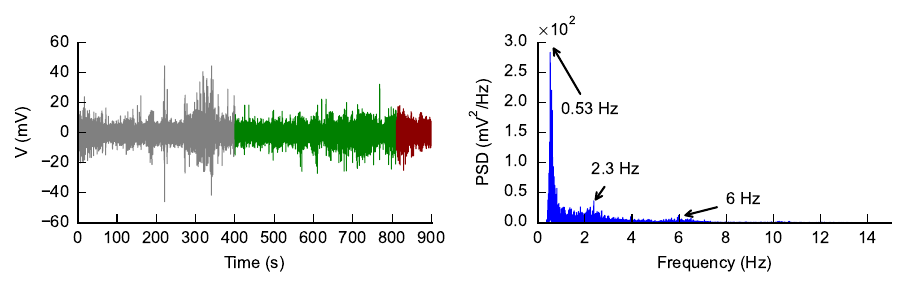}
        \subcaption{$\alpha=150^{\circ}$}
    \end{subfigure}

    \caption{Panels illustrate the temporal dynamics as well as frequency content of the Radon transformed data of the same pixel located at $d=40$ in the $(d,\alpha)$ plane for two different angles: $60^\circ$ (a) and $150^\circ$ (b). Each row shows the time series of the transformed signal on the left, along with its power spectrum density (PSD) and highlighted frequency peaks on the right. The panels show similar temporal dynamics and frequency content for the two angles, as well as similar frequency profiles to the original non transformed signal of \Cref{fig:A4_2d_t}. }
    \label{fig:A4_radon_t_freq}
\end{figure}

For temporal dynamics, \Cref{fig:A4_radon_t_freq} shows the time series and PSD
of a spatial node located in the center of the one-dimensional domain for the angles
$\alpha=60^{\circ}$ and $150^{\circ}$. 
By eye, the dynamics appears similar to those observed at pixel level in
\Cref{fig:A4_2d_t}. Indeed we see that the PSD has the same structure as observed pixel-wise with a dominant peak at around $0.5\mathrm{Hz}$ and two broad peaks
around $2\mathrm{Hz}$ and $6\mathrm{Hz}$.

We conclude that using the Radon transform to reduce data from two to one dimension preserves important features in terms of spatial and temporal frequencies. In our investigation below, we will perform data assimilation for different angles $\alpha$.

\section{Nested particle filter for the Radon transformed cortical data}\label{section:results}

\subsection{Results}
We now move to the inference of parameters and state of the discrete model
\eqref{eq:SSM_NF}, with external input parametrised as in \Cref{section:synthetic}. 

In \Cref{fig:A4_radon_x}, we observed that the signal tends to zero at the edges
of the domain, and system \cref{eq:SSM_NF} can reproduce this behaviour with a specific
choice of the observation matrix $G$. Here we select 
$G = \mathbi{T}(J,\beta)$ where $\mathbi{T}$ is a \textit{Tukey window}, also known as 
\textit{cosine-tapered window} \cite{FJHarris}, with parameters $(J,\beta)$. For a vector of size $J$ and a tapering fraction $\beta \in [0,1]$, $\mathbi{T}(J,\beta)$ is a
diagonal $J$-by-$J$ matrix acting as a mask: the multiplication $\mathbi{T}(J,\beta)v$ tapers the first
and last $\beta/2$ components of $v$ using a cosine function, and leaves the rest
intact. The diagonal of $\mathbi{T}$ is given by

\begin{equation}\label{equation:tukey}
\mathbi{T}_{jj}=\left\{ \begin{array}{lcr}
 \frac{1}{2}\left[1+\cos{\left(\pi\left(\frac{2j}{\beta(J-1)}-1\right)\right)}\right] & \mbox{for}
 & 0 \leq j < J_\ell, \\
 1 & \mbox{for} & J_\ell\leq j \leq J_r,
 \\
  \frac{1}{2}\left[1+\cos{\left(\pi\left(\frac{2j}{\beta(J-1)}-\frac{2}{\beta}+1\right)\right)}\right]& \mbox{for} & J_r<j\leq J-1,
 \end{array}\right.
\end{equation}

where $J_\ell$ and $J_r$ are the number of samples in the left and right tapered region respectively, given by 
\[
J_\ell=\left\lfloor\frac{\beta(J-1)}{2}\right\rfloor,\quad J_r=J-1-J_\ell.
\]
In our case, we fix the tapering parameter $\beta=0.5$.
The state-space model in use is then \cref{eq:SSM_NF} in which $\Psi(X,\theta)$ is
defined with a forcing $I(t;\theta,0)$ with components $\left(\Xi(r_j,t,\theta,0)\right)_{j=0}^{J-1}$ 

\[ 
  \Xi(r,t,\theta,0) = A \cos 2 \pi \bigl(\nu r - ft\bigr),
\]
and for which 
\[
G = \mathbi{T}(J,0.5),\qquad  Q = R = 0.5, \qquad \eta_k, \gamma_k \stackrel{i.i.d}{\sim} \mathcal{N}(0,\Id_J), 
\qquad k = 1,\ldots,K.
\]
We run the nested particle filter, \Cref{alg:NPF}, with $N = M = 250$ particles,  $K = 45000 $ time steps, $\delta t=0.02\mathrm{s}$, uniform priors on parameters 
\[
    (A,\nu,f)\sim \U([0,30])\otimes\U([0,1])\otimes\U([-10,10]),
\]
and Gaussian prior on the state $X_0$, as given in \eqref{eq:GP}. Note that the size $J$ of the state vector changes depending on the Radon angle $\alpha$, and {\color{black}$\delta t$ is the inverse of the sampling rate of the LFP measurements. The tuning parameters, $Q$, $R$, $\beta$ and the prior ranges  were
chosen from a set of candidates consistent with the characteristics of the observations ( amplitude and frequencies), for which we run \Cref{alg:NPF} and monitor the behaviour of variance, $RMSE$ and effective sample size.  For some synthetic experiments illustrating the effect of misspecification of $Q$, $R$ and the prior ranges, we refer the reader to \Cref{Appendix:noise,Appendix:prior}. We also reproduce the same  experiments described in this section for two additional values of the tapering fraction $\beta$ in \Cref{appendix:tukey}.}

\begin{figure}[h!]
    \centering
            
    \begin{subfigure}[b]{.3\textwidth}
    
\includegraphics[width=\linewidth]{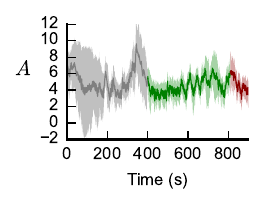}

        \subcaption{}
        \label{fig:0_I}
    \end{subfigure}
    \begin{subfigure}[b]{.3\textwidth}   \includegraphics[width=\linewidth]{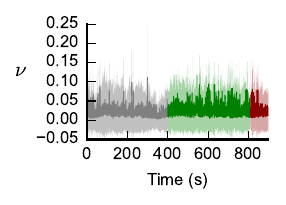}
        \subcaption{}
        \label{fig:0_j}
    \end{subfigure}
    \begin{subfigure}[b]{.3\textwidth}   \includegraphics[width=\linewidth]{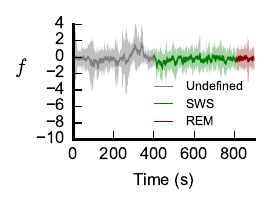}
        \subcaption{}
        \label{fig:0_f}
    \end{subfigure}

    \vfill

    \begin{subfigure}[b]{.3\textwidth}
    
\includegraphics[width=\linewidth]{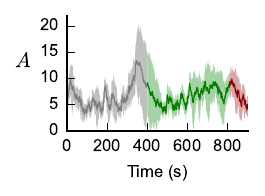}

        \subcaption{}
        \label{fig:60_I}
    \end{subfigure}
    \begin{subfigure}[b]{.3\textwidth}   \includegraphics[width=\linewidth]{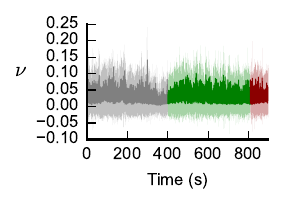}
        \subcaption{}
        \label{fig:60_j}
    \end{subfigure}
    \begin{subfigure}[b]{.3\textwidth}   \includegraphics[width=\linewidth]{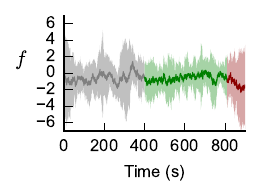}
        \subcaption{}
        \label{fig:60_f}
    \end{subfigure}

    \vfill
    \begin{subfigure}[b]{.3\textwidth}
    
\includegraphics[width=\linewidth]{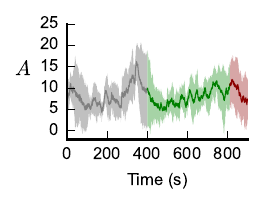}

        \subcaption{}
        \label{fig:90_I}
    \end{subfigure}
    \begin{subfigure}[b]{.3\textwidth}   \includegraphics[width=\linewidth]{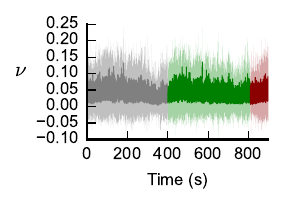}
        \subcaption{}
        \label{fig:90_j}
    \end{subfigure}
    \begin{subfigure}[b]{.3\textwidth}   \includegraphics[width=\linewidth]{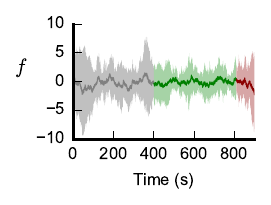}
        \subcaption{}
        \label{fig:90_f}
    \end{subfigure}

    \vfill
    \begin{subfigure}[b]{.3\textwidth}
    
\includegraphics[width=\linewidth]{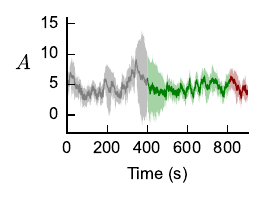}

        \subcaption{}
        \label{fig:150_I}
    \end{subfigure}
    \begin{subfigure}[b]{.3\textwidth}   \includegraphics[width=\linewidth]{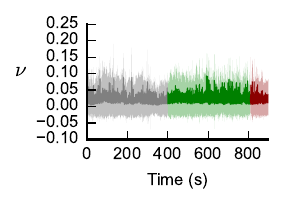}
        \subcaption{}
        \label{fig:150_j}
    \end{subfigure}
    \begin{subfigure}[b]{.3\textwidth}   \includegraphics[width=\linewidth]{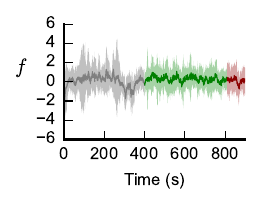}
        \subcaption{}
        \label{fig:150_f}
    \end{subfigure}

    \caption{Evolution in time of the posterior mean estimate of the input parameters for different angles $\alpha$ with $\pm 2\sigma$ band. Each panel displays the average over 10 independent runs of the nested particle filter of \Cref{alg:NPF}. Columns, from left to right, correspond to parameters of the forcing: amplitude $A$, spatial frequency $\nu$ and temporal frequency $f$. Rows, from top to bottom, correspond to Radon angles $0^{\circ}$, $60^{\circ}$,$90^{\circ}$ and $150^{\circ}$. The first 20 time steps are excluded in order to remove the burn-in period.}
  \label{fig:A4_params_angles}  
    
\end{figure}

We consider the one-dimensional observations obtained from the Radon transform with angles
$\alpha=0^{\circ},60^{\circ},90^{\circ},150^{\circ}$, as described in
\cref{sec:radon}. {\color{black} We select this subset of angles because it
spans the dominant frequency behaviours exhibited in
\cref{fig:A4_radon_x_freq_angles} without redundancy. Moreover, orthogonality of the
two pairs ensures that each selected projection angle contributes maximally
independent information about the original 2D signal.} In \Cref{fig:A4_params_angles}
we show the evolution over time of
the empirical mean estimates of the parameters $A$, $\nu$ and $f$ for the average of ten
independent runs of \Cref{alg:NPF}.
Note that the posterior mean estimates of the parameters $(A,\nu,f)$ were initialized
with $(15,0.5,0)$, however, these values are not shown on \Cref{fig:A4_params_angles}
because the first 20 time steps (burn-in) were discarded for clarity. {\color{black} We
  also show the $2\sigma$ band around the mean estimates, where $\sigma$ is the total
  standard deviation that combines the average within-run and between-run variances,
  such that

\begin{equation}\label{eq:total_var}
  \sigma^2= \frac{1}{10}\sum_{i=1}^{10}\sigma_i^2 +  \frac{1}{10}\sum_{i=1}^{10} (\mu_i-\bar{\mu})^2,
\end{equation}
where $\sigma_i^2$ and $\mu_i$ are the empirical variance and posterior mean for run $i$, and $\bar{\mu}=\frac{1}{10}\sum_{i=1}^{10} \mu_i$ is the aggregated posterior mean across runs.}

Across the four Radon angles, the posterior mean of parameter $\nu$ converges rapidly
from its initialization value of $0.5$ {\color{black} to a value between
  $0$ and $0.15$, with no major differences between sleep stages. The estimates for
  angles $60^\circ$ and $90^\circ$ (\Cref{fig:60_j,fig:90_j}) are slightly more
  elevated than those for angles $0^\circ$ and $150^\circ$
  (\Cref{fig:0_j,fig:150_j}).
 This is consistent
with the spatial scales present in the observations (\Cref{fig:A4_radon_x_freq_angles}).
 The $2\sigma$ band is tight and consistent across angles and sleep stages.}
{\color{black} This supports the
conclusion that the observed cortical patterns can be produced by a neural mass model with external forcing with spatial
frequencies  $\nu$ in the range $0-0.15\mathrm{mm}^{-1}$.}

\begin{figure}
    \centering
    \includegraphics[width=.9\linewidth]{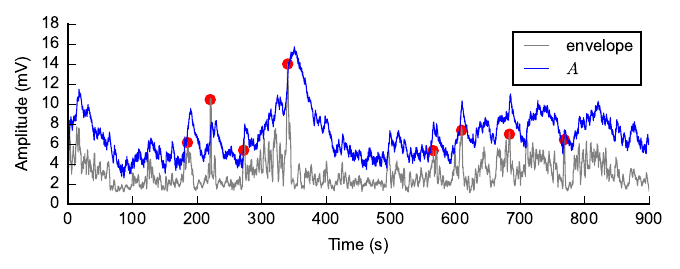}
    \caption{Posterior mean of $A$ and envelope of observed signal at $r=5.4\mathrm{mm}$ for $\alpha=60^{\circ}$.}
    \label{fig:envelope_amp}
\end{figure}

In contrast, $A$ and $f$ exhibit a more oscillatory behaviour. From an initialization
of $15$, the posterior mean estimate of $A$ settles, to a first approximation, into oscillatory
phases type $A(t) = \bar A + \tilde A(t)$. We note
that $\bar A$ changes value across Radon angles, while $\tilde A(t)$ seems to have comparable
temporal frequency. We also see abrupt changes in $A(t)$ at the onset of the slow
wave sleep (SWS) episode.

In particular, the magnitudes $\bar A$ for angles
$60^{\circ}$ and $90^{\circ}$ (\Cref{fig:60_I,fig:90_I}) are higher than those for
angles $0^{\circ}$ and $150^{\circ}$ (\Cref{fig:0_I,fig:150_I}). This is expected
since the horizontal line integrals that correspond to the radon transform for
$\alpha=90^{\circ}$ have more pixels than the vertical ones, and therefore receive on average more input. {\color{black} For $\alpha=0^\circ$, the filter starts with a wide $2\sigma$ band that narrows as the filter accumulates information, without a noticeable difference between sleep stages. For the other angles, the $2\sigma$ band widens between $300\mathrm{s}$ and $500\mathrm{s}$, which corresponds to the transition from the undefined sleep stage to SWS. For the rest of the observation window, the $2\sigma$ band remains moderate.}
Moreover, the global time trace $A(t)$ captures the dynamics
of the amplitude of the observations as illustrated in \Cref{fig:envelope_amp}. The
figure displays the posterior mean of $A(t)$
averaged over ten runs for $\alpha=60^{\circ}$ and the envelope of the
one-dimensional observations at $r=5.4\mathrm{mm}$ for the same angle. The envelope
was extracted using a moving root mean square with window length of $100$ discrete
time points corresponding to $20\mathrm{s}$. 
\Cref{fig:envelope_amp} shows how $A$ is closely linked to the dynamics of the
envelope and how the peaks of the two match, {\color{black} supporting the conclusion that our choice of the forcing $I(t;\theta,0)$ is suitable to act as a forcing term to the observed neural activity patterns.}

The red discs indicate peaks of the
envelope that exhibit gradual rise and rapid decay in the cortical signal. These
abrupt jumps cause a lag in the estimation of $A$, reflecting the efforts of the
ensemble of particles to adapt
to the new unexpected observation. This is particularly apparent for the peak around
$340\mathrm{s}$, where the envelope changes rapidly from around $18\mathrm{mV}$ to
$5\mathrm{mV}$. It takes the filter about $60\mathrm{s}$ to track this change. In
contrast, the gradual changes are tracked with little lag, see for example the
periods between $100\mathrm{s}$ and $150\mathrm{s}$, and then $450\mathrm{s}$ and
$550\mathrm{s}$. Additionally, we find that both the envelope and $A$ have a main
oscillatory component of $\tilde A(t)$ at frequency $0.002\mathrm{Hz}$. Returning to
\Cref{fig:A4_params_angles}, we observe an oscillatory behaviour that depends on the
angle for the temporal frequency parameter $f$. These oscillations appear especially
during the undefined rest and REM sleep, while $f$ appears to settle near
$0\mathrm{Hz}$ (\Cref{fig:0_f,fig:90_f,fig:150_f}) and $-0.5\mathrm{Hz}$
(\Cref{fig:60_f}) during SWS sleep. We observe also a change in sign which, in view
of the input parametrisation $\Xi(r,t;A,\nu,f,0) = A \cos 2\pi(\nu r - f t)$, indicates
a change in the direction of wave propagation from left to right.
We note also the occurrence of a burst in $A$ and $f$ simultaneously around
$340\mathrm{s}$, before the onset of slow wave sleep. {\color{black} As for
the uncertainty, estimates for angles $0^\circ$ and $150^\circ$ exhibit a moderate
and stable $2\sigma$ with few spikes across the different sleep stages. In contrast,
the uncertainty bands for $\alpha=60^\circ$ and $\alpha=90^\circ$ are initially large
for the first $100\mathrm{s}$, narrower until around $300\mathrm{s}$ when they spike
before the transition to SWS, during which they stay stable and moderate, then they
widen sharply again towards the end during REM sleep. This abrupt re-expansion of
uncertainty at the regime transition, exhibited also for parameter $A$
(\Cref{fig:60_I,fig:90_I,fig:150_I}), suggests the filter is struggling to adapt to a
changes in the dynamics.}

Next, we are interested in aggregating the estimated posterior means of $A$ and $\nu$ into time blocks and approximating their distributions within each block. This enables us to examine the global statistical structure of these estimates, particularly how they spread and cluster, rather than the instantaneous fluctuations from the time traces in \Cref{fig:A4_params_angles}.
 For this purpose, we split the estimates into blocks of $5000$ discrete time points, then approximate the density of the distribution of each block. These densities are estimated using a standard kernel density estimation method with a Gaussian kernel $K_h$ of bandwidth $h$ approximated using Silverman's rule (\cite{silverman1986density}): $h=0.9\overline{\sigma}n^{ -1/5}$, where $n=5000$ is the sample size of each block and $\overline{\sigma}$ is the sample variance.

\begin{figure}[h!]
    \centering

\begin{subfigure}[b]{.8\textwidth}
    
\includegraphics[width=\linewidth]{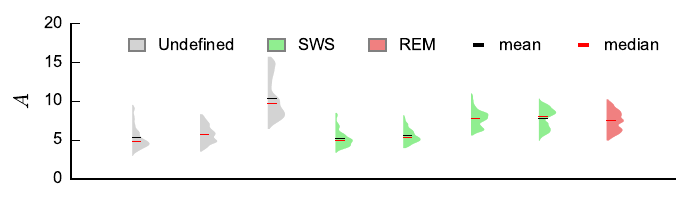}

        \label{fig:distro_60_I}
    \end{subfigure}
    
    \begin{subfigure}[b]{.8\textwidth}   \includegraphics[width=\linewidth]{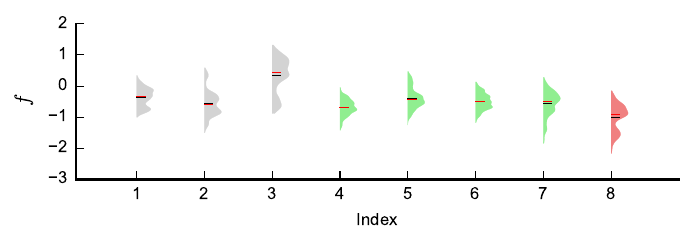}
        \label{fig:distro_60_f}
    \end{subfigure}

    \caption{Estimated density of equal blocks of the posterior mean estimates of input parameters $A$ (top) and $f$ (bottom) for angle $\alpha=60^{\circ}$. Each panels displays the kernel density estimation of blocks of $5000$ discrete time points of the average over 10 independent runs of the nested particle filter of \Cref{alg:NPF}. In each panel, index $i=1,\cdots,8$ indicates the block between $5000i$ and $5000(i+1).$}
  \label{fig:A4_distro_angles}

\end{figure}

\Cref{fig:A4_distro_angles} shows the densities of $A$ and $f$ for
$\alpha=60^{\circ}$. We observe that for both parameters $A$ and $f$, the densities are
multimodal and that the central tendencies, captured by the mean and median, as well
as the shape of the density change between blocks. This is in line with the
non-stationarity of the amplitude and temporal frequency of the observations. We
observe a sudden jump for the mean and an increase in variability in the third block
preceding the SWS between $300\mathrm{s}$ and $400\mathrm{s}$. During SWS, blocks
four and five for $A$ have similar shapes and their mean and median do not change
much (and similarly for blocks six and seven). As for $f$, blocks four to seven have
a mean around $-0.5\mathrm{Hz}$ and their shapes change smoothly. We also notice the
appearance of a tail in the lower values of $f$  (between $-1.5\mathrm{Hz}$ and
$-2\mathrm{Hz}$) for block seven, which transforms into a new cluster in that
frequency band in block eight during REM. The range of the distribution of $f$ is
consistent with the temporal frequency analysis of the observation
(\Cref{fig:A4_radon_t_freq}), which revealed the presence of two peaks around
$0.5\mathrm{Hz}$ and $2\mathrm{Hz}$. 

We conclude from the examination of the inferred input parameters that our model
(\Cref{eq:SSM_NF}) can handle the challenges posed by the LFP measurements. This consists in
modelling the forcing as a travelling space-time wave with three parameters $A$, $\nu$
and $f$ that relate to concrete features of the LFP measurements. In fact, $A$ was
shown to track the dynamics of the envelope of the observations, $\nu$  the spatial
frequencies and $f$ the temporal frequencies. The mean posterior estimates of these
parameters were consistent with the scales naturally present in the observations, and they
were able to flexibly track the non-stationary behaviour of these features. Moreover,
the transition from undefined rest state to slow wave sleep was demonstrated by the
simultaneous occurrence of a sharp spike in both $A$ and $f$. During SWS, the
parameters were shown to change gradually, and the magnitudes of the jumps, if
present, were generally smaller than  during the other sleep stages. Additionally,
the absence of significant variations by projection angle indicates that a spatially
uniform input or a superposition of waves with different and even
time‑varying—propagation directions could be a suitable model for the input when
considering the original two-dimensional LFP measurements. 

\begin{figure}[h!]
    \includegraphics[width=\linewidth]{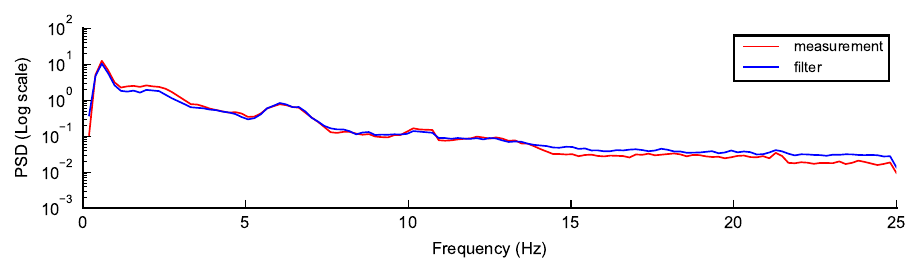}
    \caption{Temporal PSD of estimated state and observations at $x=4.2\mathrm{mm}$ for $\alpha=60^{\circ}$.}\label{fig:psd_60}
\end{figure}

Next, we shift our focus to state estimation. To assess its quality, we take as our metric the aggregated $\rmse$ as defined in \Cref{eq:rmse}. In \Cref{tab:errors} we show the averaged $\rmse$ over the
ten runs of \Cref{alg:NPF} by angle. The values reported indicate good state
reconstruction and we notice no great difference between the four angles (with only $90^{\circ}$ having a slightly higher error). In addition, we also compare the PSDs of
the observed signal and the estimated state at a single spatial node for illustration.
\Cref{fig:psd_60} shows the two PSDs, calculated using Welch's method (\cite{welch}),
on a linear-log scale at $r=4.2\mathrm{mm}$ for $\alpha=60^{\circ}$ in
\Cref{fig:psd_60}. We opted for this method instead of the single periodograms shown in the previous sections, because it provides reduced variance and smoother estimates that are easy to examine visually. We observe identical overall frequency structures, with peaks
aligned in the dominant frequency bands around $0.53\mathrm{Hz}$, $2\mathrm{Hz}$ and $6\mathrm{Hz}$.
Furthermore, since the frequency peak at $0.53\mathrm{Hz}$ corresponds to the maximum
signal energy in both cases, we can calculate the deviation of the value of this
frequency peak between the observed and estimated state across the whole spatial domain. We denote this quantity by
$\Delta f_t$, so that

\begin{equation}
    \Delta f_t:=\frac{1}{J}\sum_{j=0}^{J-1} \left|\arg \max_{\omega \in \Omega }|\F (Y(r_j,t))(\omega)|^2-\arg \max_{\omega \in \Omega }|\F (X(r_j,t))(\omega)|^2\right|,
\end{equation}
where $\F$ is the Fourier transform and $\Omega$ is the frequency domain. The average of this metric over ten independent runs is given in \Cref{tab:errors}. This shows a low deviation across the four angles, which indicates the conservation of the main frequency in time.  Our approach of joint parameter and state estimation is successful at preserving the dominant oscillatory components of the LFP measurements, which highlights its ability to accurately capture meaningful dynamics while learning the parameters of the model that was proposed to have generated them.

\begin{table}[h]
  \centering
  \begin{tabular}{ccc}
    \hline
    Angle (degree) & $\rmse$ ($\mathrm{mV}$)  & $\Delta f_t$ ($\mathrm{Hz}$) \\
    \hline
     0 & 1.6565 & 0.0080 \\
     60 & 1.9295 & 0.0409 \\
     90 & 2.4985 & 0.0222 \\
     150 & 1.5523 & 0.0331 \\
    \hline
  \end{tabular}
  \caption{Errors made in state estimation for different Radon angles}
  \label{tab:errors}
\end{table}

\section{Conclusion}\label{section:conclusion}

This work presents a proof-of-concept effort to jointly estimate the state and
parameters of a neurobiological rate network model from LFP measurements of a mouse
cortex during natural sleep. For this purpose, we used a nested particle filter on
one-dimensional observations obtained by the Radon transform of the original,
two-dimensional measurements. 

Initially, we tested this approach on synthetic data, and showed that the algorithm
correctly estimates states and learns parameters, including those that vary in time
following dynamics not explicitly included in the model. This held true for the
cortical data as well; we were able to
successfully track the observations while preserving their main frequency
characteristics through the different sleep stages and for different projection
angles in the one-dimensional reduction.

Our choice of the Amari neural field model was motivated mainly by its being one of
the simplest integro-differential equations that capture the nonlinear and nonlocal
nature of the interactions between the neurons. Moreover, there is an abundance of
results about existence and uniqueness of the solutions and their properties.
However, the model assumes a homogeneous neuronal population and parameters, which is
unrealistic for actual cortical systems. 
We also chose the simplest of discretization schemes in order to avoid unnecessary
computational overhead or the need to estimate additional parameters for more complex
schemes. Furthermore, such methods work well with acceptable precision for smooth
solutions, as is the case for the neural field. Our goal in this work was not to
achieve optimal modeling performance, but rather to demonstrate the feasibility and
potential of Bayesian estimation based on neural mass models for LFP cortical
measurements. These initial findings motivate future work on refining the state-space
models and on parameter estimation for cortical data. This includes using multiple
heterogeneous populations of excitatory-inhibitory neurons, or models that can be
more directly mapped to microscopical neural activity.







\section*{Acknowledgements}

We are grateful to Federico Stella (Donders Centre for Neuroscience, Radboud university) for providing the experimental recording used in
the study.

\section*{Declarations}


\begin{itemize}
\item Data availability. 
Imaging data used in this work have been deposited in the Donders Repository \url{https://doi.org/10.34973/2w2s-tg07}.
\item Code availability. A repository with codes will be made available. 
\end{itemize}







\begin{appendices}
\appendix
\crefalias{section}{appendix}
\section{Sensitivity analysis}\label{appendix: sensitivity}

We perform a preliminary sensitivity analysis to assess the influence of the input parameters $\theta$ on the neural field  
dynamics. We proceed in three stages. First stage: we simulate the spatio-temporal field $u(x,t)$ 
by numerically integrating the semi-discrete system \Cref{eq:semi_dis} across an ensemble of $N_s=2560$ parameter 
combinations $\theta$. This ensemble is sampled from the parameter range 
$\theta=(A,\nu,f)\in[0,10]\times[0,1]\times[0,1]$ using Saltelli’s extension 
of the Sobol sequence; this same range will be used for the priors when inferring these parameters. 
All simulations of \Cref{eq:semi_dis} are on a spatio-temporal grid $[0,10]\times[0,100]$ 
with $dr=0.02$ and $dt=0.005$, with $\tau =1$, $a=0$ and the rest of the model parameters 
are as in \Cref{table:params}. Second stage: each realization of these two-dimensional fields is reduced into
a scalar quantity of interest (QOI) via Singular Value Decomposition; we decompose
each field into modes using 
\begin{equation*}
    U=\Phi \Sigma \Psi,
\end{equation*}
where $\Phi=(\Phi_1, \cdots,\Phi_J,)$ are the spatial modes, $\Psi=(\Psi_1, \cdots,\Psi_J)$ 
are the temporal modes and $\Sigma=\textrm{diag}(\sigma_1,\cdots,\sigma_J)$ are the singular values 
with $\sigma_1\geq \sigma_2\geq \cdots$, we then extract the following scalar quantities 
of interest:
\begin{enumerate}
    \item $\eta_1$ fraction of energy of the first mode, given by 
    $\eta_1=\displaystyle\frac{\sigma_1^2}{\sum_{i=1}^{J}\sigma_i^2}$;
    \item $k_{90}$ number of modes accounting for $90\%$ of energy, calculated as 
    \[
      k_{90}=\min\Biggl\{k:
      \displaystyle\frac{\sum_{i=1}^{k}\sigma_i^2}{\sum_{i=1}^{J}\sigma_i^2}>0.9\Biggr\};
  \]
    \item $R_1$ roughness of the first spatial mode, given by $R_1=\Vert \nabla
      \Phi_1\Vert^2_2$;
    \item $T_1$ period of the first temporal mode, given by the inverse of the dominant frequency 
    $T_1=1/\arg\max_f \vert \hat{\Psi}_1\vert^2$.
\end{enumerate}

We use these different quantities of interest (rather than a single one) to examine different aspects of the model dynamics
(energy, frequencies, \dots), which yields a global picture of the model's sensitivity structure.

Third stage, we compute the first order $S1$ and total order $ST$ Sobol 
indices of the parameters $A$, $\nu$ and $f$. $S1$ and $ST$ measure the individual and total contribution of each input (parameter) to the output (QOI) variance respectively. We report their values in \Cref{tab:sobol_extended}.  

\begin{table}[h]
\centering
\begin{tabular}{l cc cc cc cc}
\toprule

& \multicolumn{2}{c}{$\eta_1$}
& \multicolumn{2}{c}{$k_{90}$}
& \multicolumn{2}{c}{$R_1$}
& \multicolumn{2}{c}{$T_1$} \\

\cmidrule(lr){2-3}
\cmidrule(lr){4-5}
\cmidrule(lr){6-7}
\cmidrule(lr){8-9}

& ST & S1
& ST & S1
& ST & S1
& ST & S1 \\

\midrule

$A$ & 0.784 & 0.405 & 0.805 & 0.1 & 0.368 & 0.076 & 0.748 & 0.355\\
$\nu$ & 0.161 & 0.063 & 0.303 & 0.063 & 0.903 & 0.493 & 0.298 & 0.082 \\
$f$ & 0.443 & 0.065 & 0.823 & 0.061 & 0.282 & 0.028 & 0.479 & 0.127\\

\bottomrule
\end{tabular}
\caption{First order (S1) and total order (ST) Sobol indices for parameters $A$,
 $\nu$ and $f$ drawn from $[0,10]\times[0,1]\times[0,1]$, and for scalar quantities of interest $\eta_1$, $k_{90}$, $R_1$ and $T_1$. }
 \label{tab:sobol_extended}
\end{table}
The three parameters exhibit moderate to high total indices $ST$ and  gap $ST-S1$ 
across all QOIs, which indicates that each parameter is influential mainly through 
its interactions with other parameters and not on its own. 
$A$ has large $ST$ and moderate $S1$ for all QOIs except $R_1$, which indicates that this parameter is very 
influential, mainly through its interactions with the other parameters.
$\nu$ primarily controls $R_1$ both directly and through interactions, and influences 
the other QOIs mainly through interactions as its $S1$ is negligible.
$f$ appears to have almost no independent effect as its $S1$ is negligible compared to $ST$ 
for all QOIs.
For the parameter inference task, this suggests that the observations can constrain combinations of the parameters, not each separately. However considering a smaller range for parameters, around the true values from the experiments of \Cref{section:synthetic} ($A=1,\; \nu=0.1, f\in[0,0.5]$), paints 
a different picture for the parameter $f$ in particular. We calculate total and first order Sobol indices for parameters in the range
$\theta=(A,\nu,f)\in[0.5,1.5]\times[0.05,0.5]\times[0,1]$, while keeping the same setting
as before. The values are reported in \Cref{tab:sobol_restricted}.
\begin{table}[h]
\centering
\begin{tabular}{l cc cc cc cc}
\toprule

& \multicolumn{2}{c}{$\eta_1$}
& \multicolumn{2}{c}{$k_{90}$}
& \multicolumn{2}{c}{$R_1$}
& \multicolumn{2}{c}{$T_1$} \\

\cmidrule(lr){2-3}
\cmidrule(lr){4-5}
\cmidrule(lr){6-7}
\cmidrule(lr){8-9}

& ST & S1
& ST & S1
& ST & S1
& ST & S1 \\

\midrule

$A$ & 0.182 &  0.066  &  0.309 & 0.018  & 0.344  & 0.018  & 0.358  & 0.060 \\
$\nu$ & 0.044 &  0.0145 &  0.162 & 0.027  & 0.574  & 0.036  &  0.712 & 0.365 \\
$f$ & 0.936 &  0.824 &  1.038 &  0.595 &  0.997 &  0.239 &  0.628 & 0.260\\

\bottomrule
\end{tabular}
\caption{First order (S1) and total order (ST) Sobol indices for parameters $A$,
 $\nu$ and $f$ drawn from $[0.5,1.5]\times[0.05,0.5]\times[0,1]$, and for scalar quantities of interest $\eta_1$, $k_{90}$, $R_1$ and $T_1$. }\label{tab:sobol_restricted}
\end{table}

In this restricted range, the three parameters exhibit different behaviours: $A$ has
negligible first order indices $S1$ across all quantities of interest, but moderate total indices $ST$. 
$\nu$ has negligible $S1$ and $ST$ except for $R_1$ where it influences mainly
through interactions, and $T_1$ for which it has high $S1$ and $ST$; finally, $f$ seems to be
dominant for all four quantities of interest through its interactions as well as
directly. This exploration of the sensitivity near the true values of the parameters
suggests that once the particles approach this restricted region in the parameters
space, inference of the parameters should be easier, since the influence of the
parameters is more direct.




\section{Robustness}\label{Appendix:robust}%
\subsection{Discretization parameters}\label{Appendix:discrete}
\begin{figure}
    \begin{subfigure}[b]{.324\linewidth} \includegraphics[width=\linewidth]{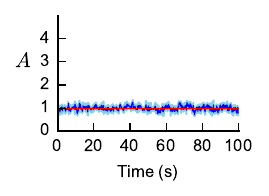}
    \subcaption{}
    \end{subfigure}
    \begin{subfigure}[b]{.324\linewidth} \includegraphics[width=\linewidth]{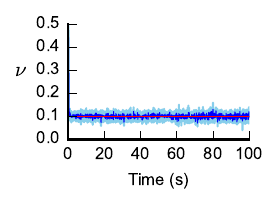}
    \subcaption{}
    \end{subfigure}
    \begin{subfigure}[b]{.324\linewidth} \includegraphics[width=\linewidth]{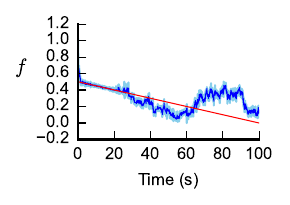}
    \subcaption{}
    \end{subfigure}
    
    \caption{$J=30,\;\Delta t=0.02$}
    \label{fig:Chirp_N5k}
\end{figure}

\begin{figure}
    \begin{subfigure}[b]{.324\linewidth} \includegraphics[width=\linewidth]{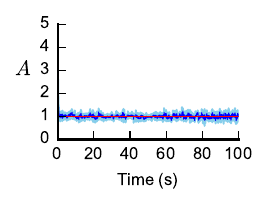}
    \subcaption{}
    \end{subfigure}
    \begin{subfigure}[b]{.324\linewidth} \includegraphics[width=\linewidth]{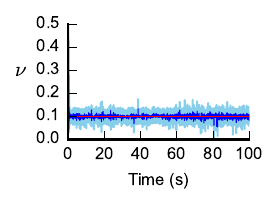}
    \subcaption{}
    \end{subfigure}
    \begin{subfigure}[b]{.324\linewidth} \includegraphics[width=\linewidth]{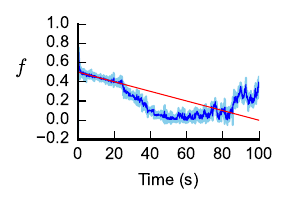}
    \subcaption{}
    \end{subfigure}
    
    \caption{$J=30,\;\Delta t=0.005$}
    \label{fig:Chirp_N20k}
\end{figure}

\begin{figure}
    \begin{subfigure}[b]{.324\linewidth} \includegraphics[width=\linewidth]{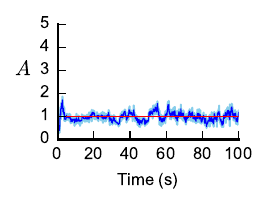}
    \subcaption{}
    \end{subfigure}
    \begin{subfigure}[b]{.324\linewidth} \includegraphics[width=\linewidth]{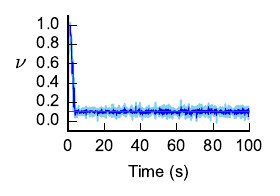}
    \subcaption{}
    \end{subfigure}
    \begin{subfigure}[b]{.324\linewidth} \includegraphics[width=\linewidth]{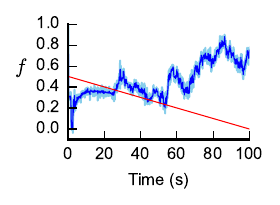}
    \subcaption{}
    \end{subfigure}
    
    \caption{$J=15,\; \Delta t=0.01$}
    \label{fig:Chirp_J15}
\end{figure}

\begin{figure}
    \begin{subfigure}[b]{.324\linewidth} \includegraphics[width=\linewidth]{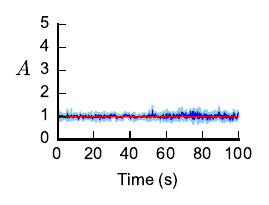}
    \subcaption{}
    \end{subfigure}
    \begin{subfigure}[b]{.324\linewidth} \includegraphics[width=\linewidth]{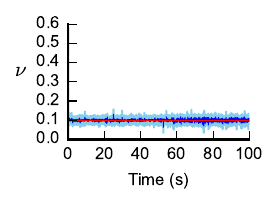}
    \subcaption{}
    \end{subfigure}
    \begin{subfigure}[b]{.324\linewidth} \includegraphics[width=\linewidth]{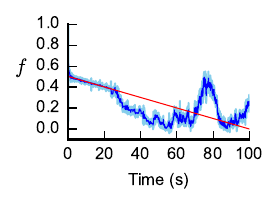}
    \subcaption{}
    \end{subfigure}
    
    \caption{$J=45,\; \Delta t=0.01$}
    \label{fig:Chirp_J45}
\end{figure}

While an extensive study of the effect of discretization schemes on inference is beyond the scope of this paper, we present some evidence that our results are robust to changes in the temporal and spatial discretization parameters $\Delta t=0.01$ and $J=30$ used in \Cref{section:synthetic}. We only require that these steps permit us to resolve the frequencies present in the observations. For this purpose, we conduct the same experiment as described in \Cref{section:synthetic} while only changing $\Delta t$ and $J$. In \Cref{fig:Chirp_N5k}, we show the effect of using a $\Delta t$ that is double the one used in \Cref{section:synthetic}. We observe that the posterior mean estimates of $A$ and $\nu$ still converge to the true value, while $f$ exhibits a similar behaviour to \Cref{fig:syn_f0} for the first $60\mathrm{s}$, after which its deviations from the truth are higher than in \Cref{fig:syn_f0}. In \Cref{fig:Chirp_N20k}, where $\Delta t$ is half the one used in \Cref{section:synthetic}, the estimates of $A$ and $\nu$ behave the same as in \Cref{fig:syn_I0,fig:syn_j0}. However, the estimate of $f$ for this smaller $\Delta t$ seems to track more closely the temporal frequency variations exhibited by the state in \Cref{fig:time_syn_cos}. \Cref{fig:Chirp_J15}, which corresponds to a smaller state dimension $J$ (a bigger spatial discretisation step $\Delta r$), shows slower convergence for $A$ and $\nu$ with more fluctuations around their true values, while the estimate for $f$ manages to remain relatively close to the truth until around $60\mathrm{s}$ after which it diverges. In \Cref{fig:Chirp_J45}, increasing $J$ (smaller $\Delta r$) yields posterior mean estimates of $A$ and $\nu$ that converge almost instantly with small and stable variance, and a behaviour similar to \Cref{fig:syn_f0} for $f$. Across these tests, the posterior mean estimates of $A$ and $\nu$ remain stable across varying and reasonable discretisation resolutions. As for $f$, the discretisation steps must be sufficiently small to ensure that the numerical integration of the evolution model captures the observed dynamics of this parameter.

\subsection{Noise intensities}\label{Appendix:noise}

\begin{figure}
    \begin{subfigure}[b]{.324\linewidth} \includegraphics[width=\linewidth]{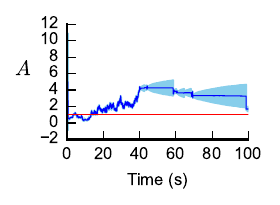}
    \subcaption{}
    \end{subfigure}
    \begin{subfigure}[b]{.324\linewidth} \includegraphics[width=\linewidth]{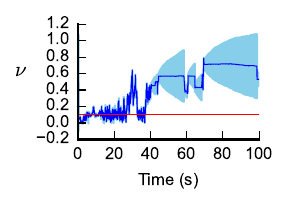}
    \subcaption{}
    \end{subfigure}
    \begin{subfigure}[b]{.324\linewidth} \includegraphics[width=\linewidth]{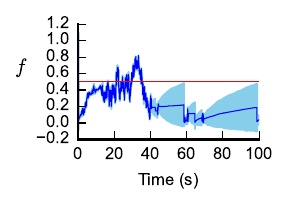}
    \subcaption{}
    \end{subfigure}
    
    \caption{$q=0.1,\ R=0.1$}
    \label{fig:q0.1_R0.1}
\end{figure}

\begin{figure}
    \begin{subfigure}[b]{.324\linewidth} \includegraphics[width=\linewidth]{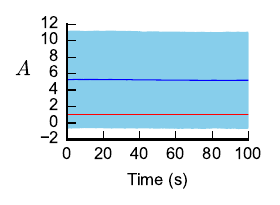}
    \subcaption{}
    \end{subfigure}
    \begin{subfigure}[b]{.324\linewidth} \includegraphics[width=\linewidth]{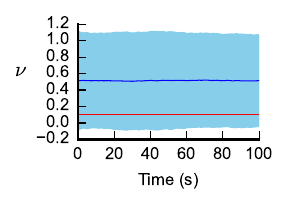}
    \subcaption{}
    \end{subfigure}
    \begin{subfigure}[b]{.324\linewidth} \includegraphics[width=\linewidth]{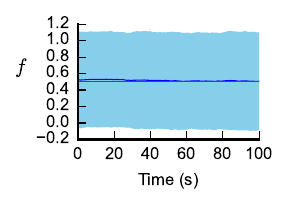}
    \subcaption{}
    \end{subfigure}
    
    \caption{$q=0.5,\ R=0.1$}
    \label{fig:q0.5_R0.1}

\end{figure}

\begin{figure}
    \begin{subfigure}[b]{.324\linewidth} \includegraphics[width=\linewidth]{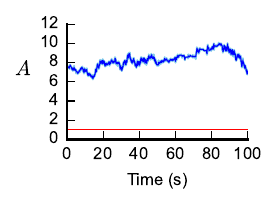}
    \subcaption{}
    \end{subfigure}
    \begin{subfigure}[b]{.324\linewidth} \includegraphics[width=\linewidth]{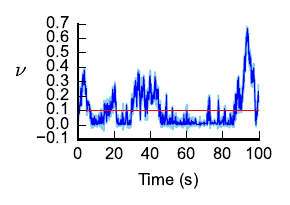}
    \subcaption{}
    \end{subfigure}
    \begin{subfigure}[b]{.324\linewidth} \includegraphics[width=\linewidth]{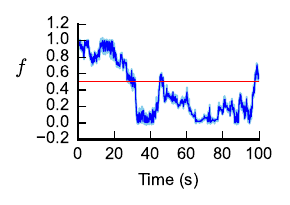}
    \subcaption{}
    \end{subfigure}
    
    \caption{$q=1,\ R=0.5$}
    \label{fig:q1_R0.5}
\end{figure}

The performance of particle filters is highly sensitive to process and observation
noise intensities, though the two play different roles. In fact, $Q$ governs the
spread of the ensemble of particles in the prediction step. If $Q$ is too small, the
particles remain tightly clustered and fail to cover the region of the space where
the true state is, which leads to an impoverished sample after resampling. If $Q$ is
too large, the ensemble will be overly spread and therefore the filter estimates lose
precision. $R$ controls the sharpness of the likelihood and therefore the weighting
of the particles. If $R$ is too small, non-negligible weights will be assigned to a
handful of particles which will then be duplicated while discarding the rest of the
ensemble through resampling, leading to a loss of diversity.
On the other hand, a very large $R$ will result in all particles receiving similar
weights regardless of their proximity to the observation, making the later non
informative. This misspecification of the noise intensities for the inner layer of the nested particle filter can directly distort the outer weights of the parameters, since these are calculated from the state weights (\Cref{alg:NPF}, line 13). Practically, this can manifest in pathological behaviours such as weight collapse or an invariant particle ensemble that does not update with new observations. To illustrate this, we run the same experiment as \Cref{section:synthetic} for different $Q$ and $R$ pairs while eliminating the model mismatch aspect. That is, we set $a^*$ to 0 for the process generating observations. 
\Cref{fig:q0.1_R0.1,fig:q0.5_R0.1,fig:q1_R0.5} show some examples of the effect of misspecified $Q$ and $R$ on the parameter inference. For \Cref{fig:q0.1_R0.1}, after an initial period where the estimates are near the true values, the filter diverges and the variance blows up. In \Cref{fig:q0.5_R0.1}, the ensemble is not moving and particle weights are not updated after each new observation. As a result, the estimates are stuck at the mean of the prior with a variance that does not shrink as new information accumulates. As for \Cref{fig:q1_R0.5}, the filter settles near the wrong values with almost zero variance, indicating a possible ensemble collapse to few particles.

\subsection{Prior range}\label{Appendix:prior}
The choice of the prior range for the parameters and the jittering kernel is also critical. A prior that does not include the true value, or has insufficient initial 
coverage near it will result in a failed recovery of the parameters. By \eqref{eq:JitKer}, the parameter 
particles are constrained to vary in the range dictated by the prior ( which is also the support of the jittering kernel), they cannot escape this range regardless of what the observations may suggest. If the true value is not included in 
the initial range, the filter will not recover it  and will produce an incorrect estimate.
This failure can manifest itself as the particles, and by consequence the posterior
mean, clustering near the boundary of the prior range closest to the true value. This can
have further consequences for the estimation of the parameters: the filter may
partially compensate for a misplaced parameter by adjusting the estimation of one or
more other parameters in an effort to make the predicted state close to the
observations.
To illustrate this, we repeat the same experiment as \Cref{section:synthetic} while changing only the prior range of 
$A$ to from $\U([0,10])$ to $\U([2,10])$ which excludes the true value $A^*=1$. \cref{fig:wrong_prior_params} shows that while we 
successfully recover $\nu$, the posterior mean of $A$ stalls near the boundary of its prior while $f$ varies to absorb the mismatch between the state and observations due to the error in $A$. A plausible choice of the prior should then be guided by physical knowledge of the parameter (known bounds, typical magnitude, etc.) and the characteristics of the observations and how they are related to the parameters.

\begin{figure}
    \begin{subfigure}[b]{.324\linewidth} \includegraphics[width=\linewidth]{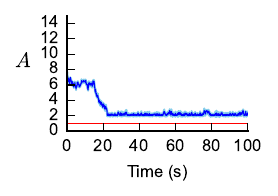}
    \subcaption{}
    \label{fig:wrong_I0}
    \end{subfigure}
    \begin{subfigure}[b]{.324\linewidth} \includegraphics[width=\linewidth]{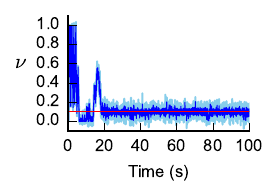}
    \subcaption{}
    \label{fig:wrong_j0}
    \end{subfigure}
    \begin{subfigure}[b]{.324\linewidth} \includegraphics[width=\linewidth]{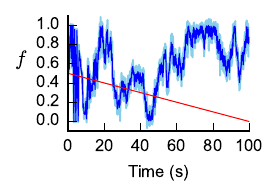}
    \subcaption{}
    \label{fig:wrong_f0}
    \end{subfigure}
    \caption{For the same numerical experiment as \cref{fig:syn_cos_params} except for changing the prior of $A$ to $\U([0,2])$ which does not include the true value $A^*=1$, we show the sequence of empirical posterior means (blue line) within $2\sigma$ band (shaded region)
    compared to their true values (red line).}
    \label{fig:wrong_prior_params}
\end{figure}

\subsection{Tukey window}\label{appendix:tukey}
\begin{figure}
    \begin{subfigure}[b]{.324\linewidth} \includegraphics[width=\linewidth]{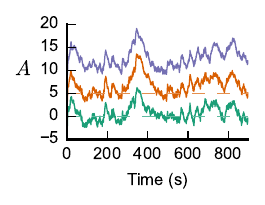}
    \subcaption{}
    \label{fig:Tukey_combined_A_off}
    \end{subfigure}
    \begin{subfigure}[b]{.324\linewidth} \includegraphics[width=\linewidth]{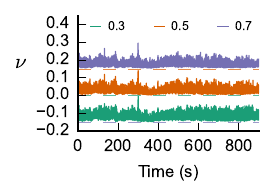}
    \subcaption{}
    \label{fig:Tukey_combined_nu_off}
    \end{subfigure}
    \begin{subfigure}[b]{.324\linewidth} \includegraphics[width=\linewidth]{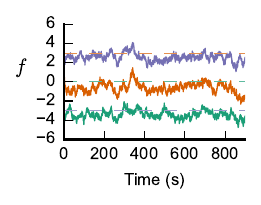}
    \subcaption{}
    \label{fig:Tukey_combined_f_off}
    \end{subfigure}

    \begin{subfigure}[b]{.324\linewidth} \includegraphics[width=\linewidth]{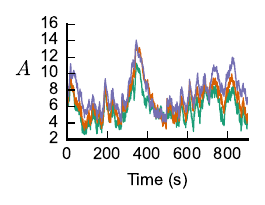}
    \subcaption{}
    \label{fig:Tukey_combined_A}
    \end{subfigure}
    \begin{subfigure}[b]{.324\linewidth} \includegraphics[width=\linewidth]{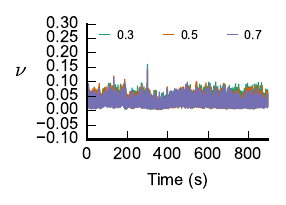}
    \subcaption{}
    \label{fig:Tukey_combined_nu}
    \end{subfigure}
    \begin{subfigure}[b]{.324\linewidth} \includegraphics[width=\linewidth]{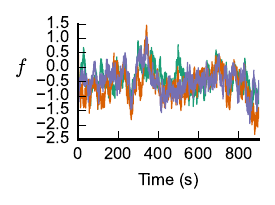}
    \subcaption{}
    \label{fig:Tukey_combined_f}
    \end{subfigure}
    \caption{Inferred parameters for $\alpha=60^\circ$ for three tapering fractions of the Tukey window. The top row shows the trajectories with an offset.}
\end{figure}

The tapering fraction $\beta$ of the Tukey window used as observation operator in the experiments of \Cref{section:results}, controls how much of the signal is preserved without modification and how much is subject to decay at the edges. 
In order to evaluate the effect of $\beta$ on the dynamics of the estimated
parameters and quality of state estimation, we run experiments in the same setting
described in \Cref{section:results} for angle $\alpha=60^\circ$ for different values of the tapering
fraction
$\beta=0.3,0.5,0.7$. \Cref{fig:Tukey_combined_A_off,fig:Tukey_combined_nu_off,fig:Tukey_combined_f_off} show the posterior means of the inferred input parameters with added/ subtracted offset to separate them, while \Cref{fig:Tukey_combined_A,fig:Tukey_combined_nu,fig:Tukey_combined_f} show them overlaid.
These figures show that the dynamics of the mean posteriors are globally conserved through the change of $\beta$, with the amplitude parameter $A$ slightly decreasing with $\beta$ and $f$ showing some slight variations between the different values of $\beta$. 
We report the aggregated $\rmse$ and frequency deviation $\Delta f_t$ for the three cases in \Cref{tab:Tukey_RMSE}. While $\Delta f_t$ remains almost unchanged, the $\rmse$ for $\beta=0.5$ and $0.7$ is lower which indicates these values are more suitable for our application.

\begin{table}[h]
\centering
\begin{tabular}{l c c}
\toprule

& RMSE & $\Delta f_t$ \\

\midrule

$\beta= 0.3$ & 2.2565 &  0.0150   \\
$\beta= 0.5$ & 2.0712 &  0.0156  \\
$\beta= 0.7$ & 2.0554 &  0.0141 \\

\bottomrule
\end{tabular}
\caption{RMSE and $\Delta f_t$ for different tapering fractions $\beta$ of the Tukey window for $\alpha=60^\circ$.}\label{tab:Tukey_RMSE}
\end{table}


\end{appendices}


\bibliographystyle{plain}
\bibliography{refs}

\end{document}